\documentstyle[12pt,epsf]{article}

\newcommand{\eq}{\begin{equation}}
\newcommand{\en}{\end{equation}}
\newcommand{\eqn}{\begin{eqnarray}}
\newcommand{\enn}{\end{eqnarray}}
\newcommand{\no}{\nonumber}
\newcommand{\nn}{\nonumber \\}
\newcommand{\hsp}{\hspace}
\newcommand{\vsp}{\vspace}
\newcommand{\sta}[2]{\stackrel{#1}{#2}}
\newcommand{\rot}[3]{\left({#1}\atop{#2}\right)_{#3}}
\newcommand{\sig}{\sigma}
\newcommand{\Sig}{\Sigma}
\newcommand{\th}{\theta}

\newcommand{\pa}{\partial}
\newcommand{\A}{\alpha}
\newcommand{\B}{\beta}
\newcommand{\ep}{\epsilon}
\newcommand{\sq}{\sqrt}
\newcommand{\fr}{\frac}
\newcommand{\pr}{\prime}
\newcommand{\ra}{\rightarrow}
\newcommand{\equ}{\equiv}

\newcommand{\ph}{\phi}

\newcommand{\La}{\Lambda}
\newcommand{\la}{\lambda}
\newcommand{\de}{\delta}
\newcommand{\De}{\Delta}

\newcommand{\vs}[1]{\vspace{#1 mm}}

\renewcommand{\thefootnote}{\fnsymbol{footnote}}
\newcommand{\ite}{\begin{itemize}}
\newcommand{\itn}{\end{itemize}}

\newcommand{\vn}{\vec{\nabla}}
\newcommand{\Exp}{{\rm e}}
\newcommand{\Nc}{{\rm N}_{\rm c}}
\newcommand{\Nf}{{\rm N}_{\rm f}}
\newcommand{\gs}{g_s}

\newcommand{\dilm}{{\rm e}^{-\phi}}
\newcommand{\Ap}{\alpha^{\pr}}
\newcommand{\ls}{{\Ap}^{\fr{1}{2}}}
\newcommand{\Laq}{\Lambda_{{\rm QCD}}}
\newcommand{\bns}{b^{{\rm NS}}}

\newcommand{\tvn}{\tilde{\nabla}}

\begin{document}
\begin{titlepage}

\begin{flushright}
hep-th/0211162
\end{flushright}

\begin{center}
{\Large \bf
Type IIB 2-form Fields and Gauge Coupling
Constant of 4D ${\cal N}$=2 super QCD}
\vs{10}

{\large
Takuhiro Kitao}\footnote{e-mail address: kitao@post.kek.jp}\\
\vs{5}
{\em Hadron-Nuclear-Quantum Field Theory Group,\\
Institute of Particle and Nuclear Studies, \\
KEK, High Energy Accelerator Research Organization,\\
1-1 Oho, Tsukuba-shi, Ibaraki-ken, 305-0801 Japan
}
\vs{10}

{\bf Abstract}
\end{center}

We study the relation between the Type IIB (NSNS and RR) 2-form fields
and the (complex) gauge coupling constant of the 4D ${\cal N}$=2 ${\rm
SU(\Nc)}$ super Yang-Mills theory with ${\rm N_f}$ fundamental
matters. We start from the analysis of the D2-brane world volume theory
with heavy $\Nc$ quarks on the $\Nf$ D6 supergravity background. After a
sequence of T- and S-dualities, we obtain the
(generalized) 2-forms in the configuration with $\Nc$ D5-branes wrapping on the
vanishing two-cycle under the influence of the background. These 2-forms
shows the same behavior as the gauge coupling constant of the 4D ${\cal
N}$=2 super QCD. The background reduces to the ${\rm Z}_{\Nf}$ orbifold
in the twelve-dimensional space-time formally realized by introducing
the two parameters as the additional space coordinates. The 10D gravity
dual is suggested as the 2D flip in this twelve-dimensional space-time.
In the case of
$\Nf= 2\Nc$, this gravity dual becomes
${\rm AdS_5 \times S^5/Z_2 } $ with D3-charge which depends
on the constant generalized NSNS 2-form. This is the result expected from 
the M-theory QCD configuration. Based on the
known exact result, we also discuss this configuration after including
the nonperturbative effect.

\end{titlepage}
\renewcommand{\thefootnote}{\arabic{footnote}}
\setcounter{footnote}{0}

\section{Introduction and Conclusions}
Recently, the generalization of ${\rm AdS_5}$/CFT$_4$ correspondence
\cite{Mald} to the non-conformal cases has been actively studied. The
first success is the discovery of the correspondence between the Type
IIB (NSNS and RR) 2-form fields and the complex (including theta
parameter) gauge coupling constant of the 4D non-conformal super
Yang-Mills (SYM) theory realized on this configuration \cite{9911096}.
After this discovery, the corresponding supergravity (SUGRA) solution
has also been constructed for ${\cal N}$=2 \cite{0011077} and ${\cal N}$=1
\cite{0002159} non-conformal SYM theories.
Especially, the case with ${\cal N}$=2 supersymmetry is 
easy to handle due to its higher supersymmetry. It is also useful 
to discuss the
theories with ${\cal N}$=1 supersymmetry or without supersymmetry. 
A lot of studies have been done in this direction 
\cite{0007018}-\cite{0205204}.

Above all, the gauge theory with the fundamental matters (fundamentals)
has attracted a lot
of interest. One reason for this is that the ratio between the number
of flavors and the rank of the gauge theory appears with the typical
coefficient in the behavior of the one-loop renomalization group (RG)
flow of the gauge coupling constant. We can compare this fact with the
behavior of the corresponding SUGRA fields.\footnote{
When we quantitatively examine AdS/CFT correspondence, we have to
compare the ratio of the two different kinds of quantity to kill
the normalization with convention dependence. 
For example, the gauge coupling
constant of 4D {\cal N}=4 SYM on D3-branes is proportional to the 
string coupling constant $\gs$, but the coefficient is the convention
dependent. In \cite{9611050}, this coefficient is fixed by 
requiring the ratio between the tension of the fundamental string 
and that of the D-string to be ${1}/{\gs}$.}
This will manifest the relation between the radial coordinate of the SUGRA solution
and the energy scale of the field theory, as originally suggested in
\cite{Mald}.

Some attempts have been made \cite{0106014}\cite{0107057} for 
this problem in the 4D ${\cal N}=2$ case.  But, as found and discussed
in \cite{0106014}, the obtained SUGRA solution does not show 
the properties required as the gravity dual corresponding to
the 4D ${\cal N}=2$ field theory.\footnote{In \cite{0107057}, 
they have suggested an interpretation
that all the string perturbation effect is included in this SUGRA
solution.} 
For example, if we can reproduce the correct RG-flow on the supergravity side,
this RG-flow will vanish 
in the special case corresponding to the 4D ${\cal N}=2$ CFT.
In this case, we can also expect that the SUGRA
solution will have 
the structure of AdS$_5$ which has the conformal symmetry. 
This is the required condition coming from AdS/CFT correspondence
and gives us the check whether the analysis is correct or not.
But, as commented in \cite{0106014}, their approach
does not satisfy this condition. 
So this problem remains unsolved and we need some modifications
of their approach to obtain the correct result.
This is the main purpose of this paper and the motivation of this study.

On the other hand, 
there is well-known successful example to make clear the structure
of the 4D field theory vacua in Type IIA string theory $-$ M-theory QCD
(MQCD) \cite{9703166}. In this model, the analysis of supersymmetric cycle
enables us to study 4D gauge field theory including nonperturbative
effect. To solve the above difficult problem, it would be the best way
to reconsider this successful model. By knowing what situation and 
how this model works, we may find the solution to the above 
mentioned problem.\footnote{
This line is also referred in \cite{0106014} as the future problem.}
In fact, it is suggested that the Type IIB configurations 
corresponding to the SUGRA solutions
are the T-dual of the MQCD configurations \cite{9811004}\cite{9811139}.
In \cite{9902210}, this is explicitly confirmed 
that a special kind of Type IIB SUGRA solution 
such as AdS$_5\times {\rm S^5/Z_2}$
\cite{9803015} is the T-dual of the Type IIA SUGRA solution 
corresponding to the MQCD configuration.

There is the well-known MQCD configuration corresponding to
the 4D ${\cal N}$=2 ${\rm SU(\Nc)}$ super Yang-Mills theory with ${\rm N_f}$
quarks in the fundamental representation. The above equivalence
by the T-duality means there is also the Type IIB configuration corresponding 
to this field theory. Therefore we can expect that we will have
the Type IIB SUGRA dual from this configuration.
This SUGRA dual will reproduce the typical RG-flow of this field theory
and have the AdS$_5$ structure for the special value ${\rm N_f}$=2${\rm \Nc }$.  
For this purpose, we need a careful treatment of this T-duality
in order to apply the knowledge of the MQCD analysis.
By this analysis, 
it will be possible to find how to modify the 
previous approach to obtain the correct result. 

In this paper, we follow this direction and study the relation between
the Type IIB 2-forms and the (complex) gauge coupling constant of the
4D ${\cal N}$=2 ${\rm SU(\Nc)}$ super Yang-Mills theory with ${\rm N_f}$
quarks in the fundamental representation. 
The outline of the strategy and results of this paper is as follows.

First, we reconsider the MQCD configuration with the $\Nf$ 
D6 background. We can find that the RG flow of the gauge coupling constant
is determined not by the simple difference of the 
coordinate of the positions for the two NS5-branes, but
by that of the newly defined coordinate.
From this observation we can point out that it will be also
true for the (T-dualized) Type IIB system; 
a newly defined field (twisted sector) is required for the gauge
coupling constant.
The use of the different type of the twisted sector
is one of the most different points from the previous attempts 
\cite{0106014}\cite{0107057}.
This gives us the clue about how to treat the problem.

For this purpose, 
we start from the analysis of one D2-brane world volume theory with
heavy $\Nc$ quarks on the $\Nf$ D6 supergravity background. This is the
T- and S-dual of the MQCD-like configuration; $\Nc$ 
semi-infinite D4-branes terminated on one NS5-brane under the
$\Nf$ D6 supergravity background. This D2-brane model is easier to
handle, so we study this configuration. In fact, at this stage we can
see that the behavior of the newly defined world volume field 
is the
same as that of the RG flow of the gauge theory. 
We can simply generalize this analysis to the case with two D2-branes. 

Here, as a by-product of this
analysis, we can also show that 
there are two kinds of definitions for the electric charge
on the D2-branes according to their relative positions to the
D6-branes. This relative difference will be related to 
the string creation (so-called Hanany-Witten
effect \cite{9611230}), but this shows that the observer on the 
D2 brane does not see the
string creation. We can also show that there are $\Nf +1$ inequivalent
BPS configurations which can not be continuously 
transformed into each other.
This corresponds to the s-rule \footnote{
When this work is being completed, we receive the paper \cite{0211020}
in which they have confirmed one aspect of s-rule that there is the maximum
of N for the continuous string charge connected between N D3-brane and
one D5-brane.}\cite{9611230}.

Next, we estimate the form of the 
background on the D3-brane which is the T-dual of the previous
D2-brane. 
Here we have to emphasize that we use the correspondence 
between the world volume scalar and
the Wilson line under the T-duality.
As a result of that, the background is
different from the D7 SUGRA solution, 
although the background is the D6 SUGRA solution in Type IIA.
This is the another most different point from the previous attempts
\cite{0106014}\cite{0107057}. Their analysis corresponds to
treating this configuration as the D3-brane on the background
of the D7 SUGRA solution. 
But, remember that one special limit for the compactified
radius is required to get the D7 SUGRA solution
from D6 SUGRA solution. We have to check whether this limit is
consistent with the condition for the realization of the field theory.
Then we can see that
it is only by our procedure that we can transfer the successful result
of the Type IIA configuration to that of the Type IIB configuration.

Then we rewrite the world volume gauge field in terms of the NSNS and RR
2-forms in Type IIB theory. After a sequence of T- and S-dualities we
obtain the generalized Type IIB 2-forms in the aimed configuration; $\Nc$ 
D5-branes wrapping on the vanishing two-cycle between
the two Kaluza-Klein (KK) monopoles under the influence of the
background. These 2-forms show the same logarithmic behavior as the RG
flow of the 4D ${\cal N}=2$ SU($\Nc$) Super QCD (SQCD) with $\Nf$
flavors. 

Here we have to note that these 2-form fields originate from 
the world volume fields corresponding to the two-dimensional
space in M-theory.
This two-dimensional space is related by the T-duality to 
the torus with the complex structure
made up of dilaton and axion in Type IIB theory \cite{9508143}.
In other words, the above 2-form fields correspond to the new
coordinates of the additional two-dimensional space for 
Type IIB theory.   
By including these two degrees of freedom as the new space coordinates,
we formally
extend our discussion to twelve-dimensional space-time.
Then we see that the background reduces to the ${\rm Z}_{\Nf}$
orbifold in this twelve-dimensional space-time.
Therefore our configuration reduces to the one embedded in 
this locally flat background.
By this procedure, we can extract or separate the gravity induced
by branes with (open string) dynamics from the background.

The another important point here
is that we can separate the another 
two-dimensional space from this twelve-dimensional
space-time. This two-dimensional space
decouples from the remaining ten-dimensional space-time. That is, what
we have done is to add the extra two-dimensional space and to pick up
the unimportant another two-dimensional space from the
twelve-dimensional space-time. This is the procedure
similar to the M-theory flip. 
On the other hand, in F-theory it is known that the
extra two-dimensional space corresponds to the space
for the dilaton and axion of Type IIB theory. 
In the context of F-theory, our procedure is
replacing the two-dimensional space for
the non-trivial dilaton and axion with another
two-dimensional space for the constant dilaton and axion.
That is, we take the frame of the (remaining) ten-dimensional space-time
in which the dilaton and axion are constant.
In this frame, the generalized 2-form (twisted sector) 
becomes the ordinary
one. Our suggestion is that this remaining 
ten-dimensional space-time would be the gravity dual of the corresponding
field theory. 
Then we can find that the configuration in the remaining ten-dimensional
space-time is qualitatively the same as that of pure SYM theory. This is
consistent with the fact that at one-loop level, the structure of pure
SYM vacua is qualitatively the same as the Coulomb branch of SQCD.

By applying the 10D SUGRA solution for pure SYM
theory \cite{0011077}, we can obtain the explicit form of 
the aimed 10D gravity dual realized as
the 2D flip in the twelve dimensions. 
Especially, in the case of $\Nf=2\Nc$, this 10D gravity dual
reduces to ${\rm AdS_5 \times S^5/Z_2 }$ 
with D3-charge which depends
on the constant generalized NSNS 2-form. This is the result expected from 
the corresponding MQCD configuration in which 
the $\Nc$ D4-branes are wrapping on only a part of the circle.  

Until this stage, we study our configuration in the region where we can
ignore the nonperturbative effect in 4D field theory. 
Based on the exact
result known purely in the field theory, 
we speculate how our configuration will be described.  We find
that the classical $\de$ function-like singularities as the source of
the D5 charge change into those of the branch cuts and there is a new
type of 'flux' \footnote{The quotation marks are added 
because the flux is originally the name for RR and NSNS 
2-form before taking the nonperturbative effect.}
which goes round between one
branch cut and another branch cut.  

This paper is organized as follows. In section 2, we reconsider
why MQCD analysis for the SU($\Nc$) SQCD with $\Nf$ fundamental matters
works well and anticipate what we should do in our configuration.
In section 3, we analyze the D2-brane world volume field theory
with heavy $\Nc$ quarks on the $\Nf$ D6-brane background.
We show that in this stage, the behavior of the scalar field
is the same as the RG flow of the SQCD. In section 4, we discuss how
the previous result is transferred to the T-dualized 
configuration. In section 5, we transform the field on the 
world volume to those of the NSNS and RR 2-forms of Type 
IIB. In section 6, by using the sequences of T and S-dualities,
we bring our all results to the aimed configuration. Then we
discuss the correspondence between the Type IIB 2-forms and
the RG flow of the gauge theory. We also discuss the 10D
gravity dual in the formal twelve dimensions.
In section 7, based on the exact result, we speculate how 
our configuration is described  when the nonperturbative effect
in 4D field theory is included.
\section{Analysis of 
MQCD Configuration Revisited }
Let us start from the analysis of the MQCD configuration \cite{9703166}.
This configuration consists of two NS5-branes and $\Nc$ D4-branes
suspended between them on the $\Nf$ D6-branes background.
These $\Nf$ D6-branes are also located between the two NS5-branes
with respect to $x^6$-direction
and we consider their positions as the origin in the directions of
$x^4$, $x^5$ and $x^6$.
We set their world volume and locations as follows;
\eqn
\begin{array}{rcccccccccl}
{\rm two\ \ NS5}: & 1 & 2 & 3 & 4 & 5 & - & - & - & - & 
\hsp{0.5cm}{\rm  at\ \ }x^6=x^{6\pm},\ \ x^7=x^8=x^9=0   \\
{\rm \Nc \ \ D4}: & 1 & 2 & 3 & - & - & 6 & - & - & - &
\hsp{0.5cm}{\rm  at\ \ }x^4=x^5=x^7=x^8=x^9=0\\
{\rm \Nf \ \ D6}: & 1 & 2 & 3 &-&-& -&7 &8 & 9 &
\hsp{0.5cm}{\rm at\ \ }x^4=x^5=x^6=0 \no
\end{array}
\enn
Let us consider the $\Nf$ D6-branes as the background of the
D6 SUGRA solution. 
The D6 solution is given by   
\eqn 
&&ds_{10}^2= {\rm H}^{-\fr{1}{2}}\Big( \eta_{\mu \nu} dx^{\mu}dx^{\nu} 
+ \sum_{i = 7,8,9} (dx^i)^2 \Big)
+  {\rm H}^{\fr{1}{2}}\Big( dx_6^2 + |dz|^2 \Big) , \nn  
&& {\rm e }^{\phi} =  
             {\rm g}_{\rm s} {\rm H}^{- \fr{3}{4}}  ,\hsp{1cm}
{\rm G}_{ab}^R = 
           -\pa_c {\rm H} \ep_{ab}^{\ \ c} ,  \label{D6sol}  \\
&& \hsp{2cm} {\rm H} \equiv 1 + \fr{\gs \ls \Nf}{2{\rm r}} ,
\hspace{2cm} {\rm r} \equiv \sqrt{|z|^2 + x_6^2},  
\hspace{2cm} z \equiv x^4 + i x^5 ,\no
\enn   
where ${\rm e }^{\phi}$ and ${\rm G}_{ab}^R$ are the dilaton and
RR 2-form field strength. 
$\ep^{abc}$ is the 3-cyclic epsilon tensor 
valued as $\ep^{456}=-\ep^{546}=\ep^{564} = +1$ and 
$\ep_{ab}^{\ \ c}= \ep_{abc}$ etc, 
and we denote the indices $\mu$, $\nu$ and the metric $\eta_{\mu \nu}$
as the indices and the flat metric of the four dimensional
space-time $x^0$,$x^1$,$x^2$, $x^3$.

Let us consider the four dimensional gauge theory on the $\Nc$ D4-branes.
This gauge theory is realized on the D4-branes after the 
dimensional reduction in the direction of $x_6$.
The renomalization group (RG) flow of the gauge coupling constant 
is given by the equation :
\eqn
\fr{1}{g_{YM}^2} &=& 
\fr{1}{\ls} \int_{x^-}^{x^+} dx_6  \Exp^{-\phi}
       \sqrt{-g_{66} {\rm det}\{ g_{\mu \nu} \} } \ \ 
       {\rm H} \nn
\hspace*{-2.5cm}&=& \fr{1}{g_s \ls} \int_{x^-}^{x^+}  dx_6 
      \Big( 1+ \fr{g_s \ls {\rm N_f} }{2\sqrt{|z|^2 + x_6^2}} \Big)
= \fr{\De x_6 }{g_s \ls} + \fr{\Nf }{2} 
 \ln 
 \left(\fr{x_6^{+} + \sqrt{|z|^2 + x_6^{2+}}}{x_6^{-} 
              + \sqrt{|z|^2 + x_6^{2-}}}\right) ,
\label{gauge}
\enn 
where the additional factor ${\rm H}$ of the 
second equation is needed in order that the 
field strength with the up-indices $F^{\mu \nu}$ is defined as 
$F^{\mu \nu} = \eta^{\mu \rho} \eta^{\nu \la} F_{\rho \la}$.

Let us consider the limit in which the quantities appeared in 
the four dimensional
field theory will remain finite. We take the limit below\footnote{
In the definition of $u$, we add the factor $\fr{1}{2\pi}$ to
simplify the equations below.}.
\eqn
 \gs \ra 0 \hsp{1cm} \Ap \ra 0 \hsp{1cm}
&&   x^6 \ra 0  \hsp{1cm} z \ra 0   \nn
\La  \equ \fr{\gs }{4 \pi {\Ap}^{\fr{1}{2}}} \ \ , \hsp{0.5cm}
&&   \ph \equ \fr{x^6}{2\pi \Ap \La} \ \ , 
   \hsp{0.5cm} u \equ \fr{z}{2\pi \Ap} \hsp{0.5cm} {\rm fixed} 
\label{limit}
\enn
Note that in this limit, the above gauge coupling constant remains finite.
Let us compare the gauge coupling constant of pure SYM with that of 
SQCD. In the case of pure SYM, 
the RG flow of the gauge coupling constant is expressed
by the distance $\De x^6$. But the different point in SQCD is the existence of 
the second term of Eq.(\ref{gauge}). Due to this term, 
the gauge coupling constant is not expressed 
only by the distance $\De x^6$ and comes to depend
on the explicit coordinates $(x^6_{\pm})$. It seems that we  
need two degrees of freedom to express the one
degree of freedom for the gauge coupling constant.
In this sense, the original coordinate $x^6$ is not 
as good as in the case of pure SYM theory.
This suggests that we should take another coordinate 
by which we can treat the
gauge coupling constant in the same way as pure SYM theory.   
Let us take the new coordinate which satisfies this requirement as
\eq
d {\tilde x_6} \equ dx_6  \Exp^{-\phi}
       \sqrt{-g_{66} {\rm det}\{ g_{\mu \nu} \} } \ \ 
       {\rm H}\hsp{1cm} {\rm  for\ \ fixed\ \ z}\ \ \ \ .
\en
By the integral with respect to $x^6$, we get 
\eq
\fr{{\tilde x_6}}{g_s \ls} = \fr{x_6 }{g_s \ls} + \fr{\Nf }{2} 
 \ln \left(\fr{x_6 + \sqrt{|z|^2 + x_6^{2}}}{2\pi \Ap \La}\right)
= \fr{\ph}{2} + \fr{\Nf }{2} \ln 
\left(\fr{\ph \La + \sqrt{|u|^2 + (\ph \La)^{2}}}{\La}\right),
\en
where we add the appropriate integral constant 
to make dimensionless. 
By using this new coordinate,
we get the behavior of the gauge coupling constant as 
$1/g_{YM}^2 = \De {\tilde x^6 }/\left({g_s \ls}\right) $.
As a result of that, we can embed all the effect of the D6-branes 
in the difference of the new coordinate $\De {\tilde x^6 }$. This
enables us to treat this field theory in the same way as the 
case on the flat background.   

Next, let us consider the theta parameter of the field theory.  
The theta parameter is determined by the distance between the two
NS5-branes in the direction of $x^{10}$. The Type IIA configuration is
delocalized in this direction, but we can  keep the distance 
between them as the phase. 
So the information about this relative distance remains even in
Type IIA theory.\footnote{There is another contribution coming from  
the RR 1-form $C_{6}^{\rm R}$.
We can set the RR
1-form $C_{6}^{\rm R}$ as a constant, while keeping non-trivial
$C_{4}^{\rm R}$ and $C_{5}^{\rm R}$ 
which give the RR 2-form field strength in the
D6 SUGRA solution (\ref{D6sol}). We set this constant $C_{6}^{\rm R}$
to be zero below.}  
As a result of that, we obtain the relation between theta parameter
and the distance of $x^{10}$ direction, as 
$\th = \De x^{10}$, as suggested by \cite{9703166}. 
For later convenience, we define another new coordinate $\chi$ as
$\chi \equ 2 x^{10}/\left(\gs \ls\right)$.

Putting together ${\tilde x_6 }$ and $x_{10} $, we define a new 
complex coordinate $y$ as 
\eq
\ln y \equ \fr{{\tilde x^6 } + i x^{10}}{\gs \ls}
=\fr{{\tilde \ph} + i \chi}{2}= \fr{1}{2}\left\{\ph + \Nf \ln 
\left(\fr{\ph \La + \sqrt{|u|^2 + (\ph \La)^{2}}}{\La}\right) + i\chi\right\} ,
\en
and using this new coordinate, we can express the complex gauge coupling
constant in the simple form as 
\eq
\fr{1}{g_{{\rm YM}}^2} + i \th 
=\fr{\De {\tilde x^6} + i \De x^{10}}{\gs \ls}
=\fr{\De{\tilde \ph} + i\De \chi}{2}
= \ln y_+ - \ln y_- ,
\en 
where $\ln y_{\pm}$ indicate the value of $\ln y$ at $x^6 = x_{\pm}^6$.
This is the same form as that of pure SYM theory originally suggested
in \cite{9703166}. 
Therefore we can expect that 
\vsp{0.3cm}

{\it{The direct (supergravity) effect of the D6-branes 
on the four dimensional field theory 
disappears by the coordinate 
transformation $x_6 \ra {\tilde x_6} $ 
and the background reduces to the same 
as pure SYM theory in appearance.
The effect of the D6-branes is implicitly included in 
$ \De{\tilde x_6}$ and $\De x_{10}$. 
}}
\vsp{0.5cm}

In fact this new coordinate is one of the two holomorphic
coordinates of (multi-)Taub-NUT space \cite{9707258}.
The above claim is consistent with the fact that the supersymmetric
cycle written by this new coordinate reproduce the correct
Seiberg-Witten curve for 4D N=2 SQCD. 

The above observation is important when we consider
the T-dualized system.  When we take the T-duality in the
direction of $x_6$, $\De x_6$ will change into the NSNS 2-form field
$b^{{\rm NS}}$ coming from the twisted sector on the orbifold.  
In the case of pure SYM theory coupled with the general background, 
the gauge coupling constant\footnote{
Here we use the term, 'gauge coupling constant' in a broad sense 
including the interaction with the background (dilaton). 
In \cite{9603167}, the
quiver gauge theories are discussed and pure SYM theory is the special
case of them.} is known to be written as $\Exp^{-\ph} b^{{\rm NS}}$
in combination with dilaton $\Exp^{\ph}$ \cite{9603167}.

In the previous attempts \cite{0106014}\cite{0107057},
they have used this formula 
and interpreted as the
holographic dual of the gauge coupling constant.
This leads to complicated and abnormal behavior of the twisted sector
because of the non-trivial dilaton with logarithmic 
behavior.

But the 
above MQCD analysis casts some doubt on 
the applicability of this formula to this T-dualized (Type IIB) model.
This analysis indicates that we have to take an appropriate 'coordinate'
instead of $b^{{\rm NS}}$. Then the nontrivial $\Exp^{-\ph}$ is absorbed in
this 'coordinate' and gives no direct contribution to the behavior of the
gauge coupling constant in appearance.  In the following sections, we will
discuss these matters by starting 
to study the behavior $\De x_6$ in the simpler case.

\section{D2-branes and Strings on D6 Background } 
\subsection{Single D2-brane with Heavy Quarks on $\Nf$ D6 Background } 
Let us consider one D2-brane on the background of the $\Nf $ D6-branes.
This is the T- and S-dual of a part of the MQCD configuration; 
one NS5-brane on the
$\Nf$ D6 supergravity background. 
So it is useful to study the behavior of the D2-brane world volume 
for our investigation of the 4D RG-flow.

In the same way as the previous section, 
the world volume of the $\Nf$ D6-branes spans $x^0, x^1,
x^2, x^3, x^7, x^8, x^9$ and they are located at $x^4 = x^5 =x^6=0$.
We consider these D6-branes as the background described by the
supergravity solution in the previous section. 
What happens if we put one D2-brane in this background ?
Let us consider the field theory on the D2-brane 
whose world volume spans $x^0, x^4$ and
$x^5$. 
This D2-brane is located at $x^7=x^8=x^9=0$, but delocalized in the 
directions of $x^1, x^2$ and $x^3$.

This type of the world volume theory (so-called Dp-D(8-p) system)
has been studied in various papers especially in the
context of the string creation or the baryon vertex 
as wrapping D-branes \cite{9807179}-\cite{9712187}, \cite{0211020}.  
In the (non-wrapping) D2-D6 system such as our model, 
it is known to be inappropriate
for the analysis of string creation. This is because 
the asymptotic behavior is bad due to
its fewer dimensional world volume.
In \cite{9905118}, this D2-D6 system is discussed with the special care
as the exceptional case.\footnote{
The intensive study on the other Dp-D(8-p) systems is 
also done in \cite{9905118}.} 
Their analysis in the asymptotically flat background depends on the
additional continuous parameter $\nu$. In their discussion, 
this parameter has the origin in the partially wrapping
D2-brane in the near-horizon limit of the background. On the other hand, 
the physical system must be realized only in the case with the special
value of it; otherwise 
the RG flow of the gauge coupling constant in our problem 
does not appear with the typical coefficient.
So we need to check this point and show that it is true.
As seen in the following, 
we can also show that in general there are $\Nf +1$ 
inequivalent BPS configurations. This is the proof of the s-rule from
the world volume soliton. This gives the quantization of their 
parameter $\nu$. 

In addition to that, we want to know what will happen in
the T-dualized version of the MQCD configuration.
This D2-D6 system is the only configuration
of the Dp-D(8-p) systems which has direct analogy with MQCD.
So it is useful and necessary to treat this
system purely in the Type IIA language, without mechanically using the
analysis of the supersymmetric cycle in M-theory.

Let us study the D2-brane action on this background. The D2-brane can be
described by the Born-Infeld action as
\eq
S_{D2} = - T_{D2} \int d^3 \sig \dilm 
  \sq{-{\rm det} \left(g_{M N } \pa_{\A} X^{M} \pa_{\B} X^{N} 
   + 2\pi \Ap F_{\A \B} \right)} + \fr{1}{2\pi \gs \ls }
              \int G^R_{(2)} \wedge A_{(1)} , 
\label{BI}
\en
where $T_{D2}$ is the D2-brane tension $T_{D2}= (4\pi^2 \gs
{\Ap}^{\fr{3}{2}} )^{-1}$. We also denote the world volume coordinates of
the D2-brane as $\sig_{\A}=\{\sig_0,\sig_4,\sig_5 \}$ and $M$, $N$ run
all the ten dimension indices. Note that in the above equation, we set
the Chern-Simons term as the form in which the RR 2-form field strength
appears instead of RR 1-form gauge field. This is because in the
T-dual (T$_6$-dual) of this D2-D6 system, 
the anomaly cancellation requires that the Chern-Simons term on
the D3-brane has the RR 1-form field strength, not RR 0-form gauge
field \cite{9605033}. 
In the case of other ordinary backgrounds, the two kinds of
form are the same up to a total derivative which has no physical
meaning. But in the case such as this configuration, this total
derivative gives the additional anomaly, which makes the total anomaly of
this system zero. So the above form of Chern-Simons term will be the 
correct form.

After taking the static gauge $\sig_0 =t$ and
$\{ \sig_4,\sig_5 \} = \{x_4, x_5\}$, and assuming 
that only $X^{6}=X^{6}(x_4, x_5)$ and $A_0 =A_0(x_4, x_5)$ are
the nontrivial fields,
let us take
the limit (\ref{limit}). Then the above action reduces to\footnote{
Our analysis is limited within ${\it O}(\La^2)$, but the result is the
same if we start from the Born-Infeld action. (see appendix)} 
\eqn
S_{D2} &=& - \fr{\La}{4\pi} \int dt du^4 du^5 {\rm H}
\Bigg[\fr{1}{\La^2} + \fr{1}{2} \left|\vn_u \ph \right|^2 
  -  \fr{1}{2}\left|\vn_u a \right|^2 + O (\La^{2})\Bigg] \nn
&& \hspace*{2cm} + \fr{\La^2}{4\pi}\int dt du^4 du^5 
          \fr{\Nf}{R^2} \left(\fr{\ph}{R} - 
               \vn_u \ph \cdot \fr{\vec{u}}{R}\right)a .
\label{D2action}
\enn
Here we use the convention in which the vector $\vec{u}$
indicates the two-dimensional vector in the space $(u^4, u^5)$
and $\vn_u$ indicates the derivative with respect to
$\vec{ u}$. We also denote $a$ as the gauge field 
$a= A_0/\La $ and 
$R$ as the rescaled length of $r$ in the previous section such as
$R \equ \sq{|u|^2 + (\La \ph)^2}$. Note that this 'length' contains the
field $\ph$ which has non-trivial dependence on $(u_4,u_5)$. By using this
$R$, ${\rm H}$ can be written as ${\rm H} = 1 + {(\Nf \La)}/{R}$.

Let us add the source with $\pm \Nc$ electric charge
to the above action. In the context of the string theory,
this means that we add the semi-infinite 
$\Nc$ fundamental strings (F1) which are terminated on the D2-brane.
The signs of $\pm \Nc$ depend on which side of the D3-brane they are
terminated.
We consider the source which is delocalized
in the direction of $x^1 , x^2$ and $ x^3 $ in order that our configuration
has the isometry in these directions.

Naively, it seems that
it is enough to add the source term to the previous action such as 
\eqn
&&\De S = \pm \Nc \int d \sig^3 A_0 \de (x_4) \de (x_5) = 
\pm \Nc \La \int dt du_4 du_5 \left\{a \vn_u 
        \left(\fr{\vec{u}}{2\pi |u|^2}\right)\right\}  . \no
\enn 
But the analysis of \cite{9708147} suggests that 
we also need the source term for $X^6$ (or $\ph$) in addition
to the above source term for $A_0$ (or $a$). 
This additional source term makes the equation of motion 
consistent with the BPS condition\footnote{Their analysis is limited 
in the region which is far from the source, so this source term
does not appear explicitly in their discussion.}. 
By their results,  we can determine the form and the coefficient of 
the additional source. 
This correct source term will be 
\eqn
&&\hspace*{-0.5cm}\De S = \pm \Nc \int d \sig^3 \left\{A_0 
               + \left(2 \pi \Ap \right)^{-1} X^6 \right\}
                       \de (x_4) \de (x_5) = 
    \pm \Nc \La \int dt du_4 du_5 \left\{\left(a + \ph\right) \vn_u 
        \left(\fr{\vec{u}}{2\pi |u|^2}\right)\right\}.  \no
\enn 
From the action $S_{D2} + \De S $, we can see
the constraint for $a$
(Gauss-Low), 
\eqn 
\vn_u \left({\rm H}\vn_u a\right) &=& \fr{\Nf \La}{R^3}
\left(\ph - \vn_u \ph \cdot \vec{u} \right) 
\pm 4\pi \Nc \de (u^4)\de (u^5)  \nn
&=& \vn_u \left(-\fr{\Nf \La
\ph}{R} \fr{\vec{u}}{|u|^2} + \left\{{\rm sign} (\ph_0) \Nf \pm 
2\Nc  
\right\}\fr{\vec{u}}{|u|^2}
\right), 
\label{Gauss}
\enn 
where $\ph_0$ is the value of $\ph$ at $u=0$. The term
which include the ${\rm sign} (\ph_0)$ is needed to kill the delta
function coming from the $\vn_u \cdot \fr{\vec{u}}{|u|^2}$ of the first
term. 

When the D6-branes are located apart in the direction
of $x^6$, the last term in the above equation is generalized to
\eqn
&&\vn_u \left(-\sum_{k=1}^{\Nf} \fr{\La
(\ph -\ph_k)}{R_k} \fr{\vec{u}}{|u|^2} + 
 \left\{\sum_{k=1}^{\Nf} {\rm sign} (\ph_0 - \ph_k)  \pm 
2\Nc  
\right\}\fr{\vec{u}}{|u|^2}
\right),\nn
&& \hspace*{2cm}R_k^2 = |u|^2 + \La^2 |\ph -\ph_k|^2
\hsp{1cm} (\ph_1 < \ph_2 < \cdots < \ph_{\Nf})
\label{Gaussg}
\enn
where $\La \ph_k$ indicates
the position of each D6-brane. So there are $\Nf +1$ choices
about $\{{\rm sign} (\ph_0 - \ph_k)\}$
depending on the value of $\ph_0$. As we will see, this leads to the 
$\Nf +1$ inequivalent BPS configurations. 

The appearance of this sign term is the significant
difference between the two choices about the form
of the Chern-Simons term $-$ which of 
the RR gauge field and the U(1) gauge field 
we should keep as the gauge field, not the field strength.
If we start by keeping the RR gauge field in the form of the
gauge field, we will not have this ${\rm sign} (\ph_0)$ term. 
 
From the above equation, we can see the 
relation between $a$ and $\ph$, 
\eq 
{\rm H}\vn_u a = -\fr{\Nf \La \ph}{R} \fr{\vec{u}}{|u|^2} +
\left\{{\rm sign} (\ph_0) \Nf \pm 2\Nc  
\right\}\fr{\vec{u}}{|u|^2} .  
\label{ph_a} 
\en  
Using this relation,
we can obtain the static Hamiltonian,  
\eqn 
{\cal H} &=& \fr{1}{4\pi \La} 
\int (d{\vec u})^2 {\rm H} \left[1 + \fr{\La^2}{2} 
   \left|\vn_u \ph \right|^2 
    + \fr{\La^2 {\rm H}^{-2}}{2 |u|^2} 
          \left(\fr{\Nf \La \ph}{R} 
              - \Nf {\rm sign}(\ph_0) \mp 2\Nc\right)^2\right] \nn 
&& \hspace*{1cm} \mp \int (d{\vec u})^2 \Nc \La \ph \de ({\vec u})^2
+ \fr{\La}{4\pi } \oint
a \left(\fr{\Nf \La \ph}{R} - {\rm sign} (\ph_0) \Nf 
                     \mp 2\Nc\right) \fr{1}{|u|} .
\no
\enn  
The last term is the boundary term on the
circumference. When we set the radius of this circumference infinite,
it gives the non-zero contribution with the factor 
$\La(\mp \Nc - {\rm sign} (\ph_0) \Nf/2)$
to the static Hamiltonian.
In the case of $\Nc =0$, this is the same as the contribution coming 
from the R-sector in the open string
one-loop amplitude. 

Next, let us consider the equation of motion. We can easily obtain
\eqn 
\vn_u \left({\rm H}\vn_u \ph \right) 
&=& \fr{\Nf \La {\rm H}^{-1}}{R^3}
         \left\{-\ph - {\rm sign} (\ph_0) \Nf \mp 2\Nc\right\} \nn
&& \hspace*{0.5cm} + \fr{1}{2} \fr{\pa {\rm H}}{\pa \ph}
    \left[ \left(\vn_u \ph \right)^2 - \fr{{\rm H}^{-2}}{|u|^2} 
       \left(-\fr{\Nf \La \ph}{R} 
           +  \Nf {\rm sign} (\ph_0) \pm 2\Nc \right)^2\right]   
\mp 4\pi \Nc\de^2({\vec u})    \nn
&& \hspace*{-1.5cm}=  - \fr{\Nf \La }{R^3} 
     \left\{\ph + \left(\vn_u a \cdot \vec{u}\right) 
         \right\} + \fr{1}{2} \fr{\pa {\rm H}}{\pa \ph}
            \Big[\left(\vn_u \ph \right)^2 
                 - \left(\vn_u a \right)^2\Big]  
\mp 4\pi \Nc  \de^2({\vec u})  .
\label{eqmD6D2}
\enn 
To get the last equation, we have used the relation (\ref{ph_a}).
From this form, we can see that if there is the additional
relation $\vn a = - \vn \ph$,
the above equation will be satisfied by the Gauss-Low (\ref{Gauss}). 
So we can expect that
this is the 'almost' BPS condition.\footnote{The same form of
the 'almost' BPS condition have been appeared in the 
papers \cite{9708147}, \cite{9807179}-\cite{0104082} 
in which the world volume 
soliton or the string creation  
is discussed.} 
As a result of that, we obtain the equation to determine the behavior
of $\ph$ ,
\eq 
{\rm H}\vn_u \ph = \fr{\Nf \La \ph}{R} \fr{\vec{u}}{|u|^2} -
\left\{{\rm sign} (\ph_0) \Nf \pm 2\Nc\right\} \fr{\vec{u}}{|u|^2} .  
\label{ph_eq} 
\en 
Using the good 'coordinate' ${\tilde \ph}$ which corresponds to the
new coordinate ${\tilde x^6}/(2\pi \Ap \La)$ in section 2, 
the above equation gives us the solution :
\eq 
{\tilde \ph} \equ 
\ph +  \Nf \ln \left(\fr{R + \La \ph }{\La}\right) 
= -\left\{\pm 2 \Nc 
          -\Nf    +{\rm sign} (\ph_0) \Nf\right\}
 \ln \left(\fr{|u|}{\La}\right) +    {\rm const.}
\label{ph_solF1}
\en

We can generalize the above result to the case in which the D6-branes
are located apart in the direction of $x^6$ as
\eq 
{\tilde \ph} \equ 
\ph +  \sum_{k=1}^{\Nf} \ln \left(\fr{R_k + \La (\ph -\ph_k) }{\La}\right) 
= -\left\{\pm 2 \Nc - \Nf
          +\sum_{k=1}^{\Nf}{\rm sign} (\ph_0- \ph_k) \right\}
 \ln \left(\fr{|u|}{\La}\right) +    {\rm const.}\hsp{1cm}.
\label{ph_solF1g}
\en
This shows that there are $\Nf +1$ inequivalent BPS configurations
depending on the value of $\ph_0$. This corresponds to
the s-rule \cite{9611230}.\footnote{
In ref.\cite{0211020}, they have confirmed that there is
the maximum of $N$ for the continuous parameter $\nu$ 
corresponding to the string charge in the D3-D5 system. This is one aspect
of the s-rule, while the original s-rule \cite{9611230} is the
statement that there are $N+1$ inequivalent BPS configurations in this
model. Our result shows the quantization of this parameter $\nu$ 
from their point of view.}
These BPS configurations can not be continuously 
transformed into each other.

In addition to the bulk equation of motion, we have to consider the
variation of the boundary term. 
We obtain at infinity,
\eq 
\de \ph \left(
\vn_u \ph \cdot \vec{u}\right)- \de a \left(
\pm 2\Nc + {\rm sign} (\ph_0) \Nf \right) 
=0 .
\no
\en
This shows us that $\ph$ never becomes a constant even at
infinity.  
In fact we obtain the logarithmic flow of $\ph$
from Eq.(\ref{ph_solF1}) 
\eq 
\ph \sim  - \left(\pm 2\Nc + {\rm sign} (\ph_0) \Nf \right)\ln|u|
\label{inf_ph}
\en
This behavior is consistent with the above boundary condition. 

\begin{figure}
\epsfysize=3.5cm 
\centerline{\epsfbox{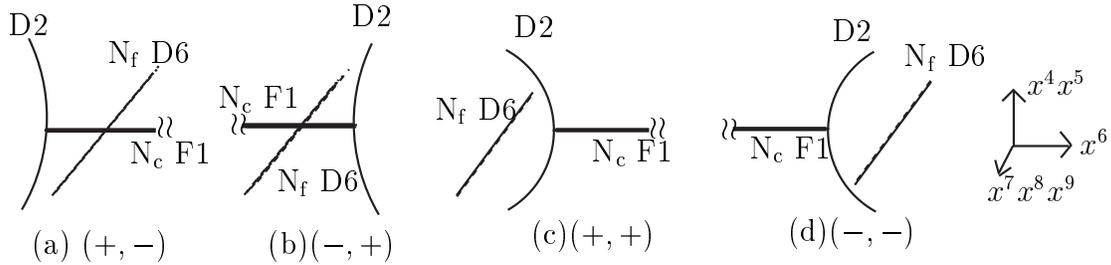}}
\caption{{\small The rough sketch for the configuration 
according to the combinations of 
(sign$\{\pm 2\Nc\}$,sign$\{\ph_0\}$)}}
\label{D2pm}
\end{figure}

In the following sections, we consider the case of $2\Nc\geq \Nf$
which corresponds to the asymptotic free or conformal 
gauge theory as we will 
see.\footnote{In the case of $\Nf > 2\Nc$, the most of our method
can be applicable except in the Ultra Violet region.} 
In this case, we can make the rough sketch Fig.\ref{D2pm} for the
configuration according to the sign of $\pm2\Nc$ and ${\rm sign}(\ph_0)$
in Eq.(\ref{ph_solF1}).
\subsection{Fundamental String Charge and Eleventh Dimension}
It is well known that the electric charge on the D-brane corresponds
to the fundamental string charge.  
In this section, we will discuss the relation between the two kinds of 
charge in our model to
connect the gauge field to the eleventh dimension.
First, we can rewrite the Gauss-Law (\ref{Gauss}) as 
\eq
\vn_u \left\{{\rm H}\vn_u a
   + \Nf\left(\fr{\La \ph}{R}  
   - {\rm sign} (\ph_0)\right)\fr{{\vec u}}{|u|^2}\right\}
=        \pm  4\pi \Nc \de (u^4) \de (u^5).
\en  
Remember that from the action (\ref{D2action}), the electric charge 
${\rm Q_E}$ is given by the integral of the left hand side as     
\eq
{\rm Q_E} = \fr{1}{4\pi} \oint *\left\{{\rm H}\vn_u a
   + \Nf\left(\fr{\La \ph}{R}  
   - {\rm sign} (\ph_0)\right)\fr{{\vec u}}{|u|^2}\right\}
 =  \pm \Nc,
\en
where $*$ means the dual in the 2D space and this integral
is calculated over the circle at the fixed $|u|$.
The right hand side of this equation indicates that
only the explicit external $\pm \Nc$ F1 source is 
the total electric charge. This is the consistent with the fact
that in our model, there is only $\pm \Nc$ F1 source 
from the starting point.
But what is the meaning of the left hand side ?
The second term is the Witten effect coming from
the Chern-Simons term. The above result shows that 
this term gives the additional
induced charge, but is canceled by 
the nontrivial contribution from ${\rm H}\vn_u a$. 
In order to give this additional contribution, the gauge field $a$ 
turns out to show the nontrivial
behavior. This makes $\ph$ nontrivial because $a$ and $\ph$ are
related by supersymmetry ('almost' BPS condition).
This is the dielectric effect
and the similar effect to Myer's effect \cite{9910053} 
which happens in another supersymmetric configuration
like D6(123789)-D2(89) system.\footnote{
By using a D-brane wrapped on a sphere, the interpretation
as Myer's effect is also given in \cite{0104082}.} 
The definition of the electric current is different according
to the sign of $\ph_0$.       
This difference 
produces the interpretation of the string creation or Hanany-Witten
effect. Note that the observer on the D2-brane never sees such
string creation because there is only external 
$\pm \Nc$ electric charge on the D2-brane.
But this relative difference is important 
for the whole system and we need the definition of the current 
which is applicable for the whole system.
  
So let us take the current of $\ph_0 > 0$ 
as the standard.
Then we define the 'dual' field as 
\eq 
\tvn_u \chi \equ {\rm H}\vn_u a + \Nf\left(\fr{\La \ph}{R}  
   - 1\right)\fr{{\vec u}}{|u|^2} \hsp{2cm}
\tvn_u \equ  \rot{-\pa_5^u}{\pa_4^u}{}
\label{chi}
\en
By this definition, we can change the dynamical variable from 
$(a,\ph)$ to $(\chi,\ph  )$.
This $\chi$ can measure the difference of 
the electric charge, namely the fundamental
string charge. This means that $\chi$ can also measure the
distance of the $x^{10}$
direction and that we can identify this $\chi$ as that
appeared in section 2.

The classical solution for $\chi$ can be obtained from its definition
and the Gauss-Low as 
\eq
\chi = \Big[\mp 2 \Nc + \Nf \left\{1- {\rm sign} (\ph_0)\right\}\Big] 
\varphi
+ {\rm const.}
\hsp{0.5cm},
\label{chi}
\en
where we define $\varphi$ by $u=u^4 + i u^5 = |u| \Exp^{i\varphi}$.
By combining this solution with that of $\ph$, we 
obtain the solution expressed in the complex 'coordinate' 
$y$ appeared in section 2 as
\eqn 
 \ln y &=&  \fr{{\tilde \ph} + i \chi}{2} 
 = \fr{1}{2}\left\{\ph + i\chi + \Nf \ln \left(\fr{R + \La
 \ph}{\La}\right)\right\} \nn
&=&\mp \Nc \ln \left(\fr{u}{\La}\right) +
      \fr{\Nf}{2} \left\{1-{\rm sign} (\ph_0)\right\} 
            \ln \left(\fr{u}{\La}\right)
          + {\rm const.}
\label{phchi_sol} 
\enn
In addition to $y$, let us define another complex variable $w$,
\eq
w \equ \Exp^{-(\ph + i \chi)/2} 
          \left(\fr{u}{|u|}\right)^{\Nf}  
              \left(\fr{-\ph \La + R}{\La}\right)^{\Nf/2} .
\en
Then we can express the above solution by using these complex
variables as
\eqn
y &=& \left(\fr{u}{\La}\right)^{\mp \Nc} \times {\rm const.} 
\hspace{3cm} \ph_0 > 0 \nn
w &=& \left(\fr{u}{\La}\right)^{\pm \Nc}   
\times {\rm const.} \hspace{3cm} \ph_0 < 0 . \no
\enn
This expresses the holomorphic embedding in
the four dimensional space with $A_{\Nf -1}$ singularity, $y w
=(u/\La)^{N_f}$. This is the expected result from the analysis when we lift
our model to M-theory and study the supersymmetric cycle of the M2-brane
on the multi Taub-NUT background with coincident $\Nf$ monopoles. 

Note that until now, we have considered the case
in which the external $\Nc$ strings have infinite length. This means
that the source with the $\Nc$ charges is too heavy to have the
dynamics.  This is why our analysis in Type IIA theory agrees with that
of M-theory. If these strings are not infinite, that is, 
the source is not so heavy, there will be some
dynamical effect as it happens in the case of MQCD. Therefore 
the analysis in
Type IIA theory is limited within the approximation in which we can
ignore this effect.

\subsection{Two D2-branes with Heavy Quarks 
on D6 Background }
\begin{figure}
\epsfysize=3.0cm 
\centerline{\epsfbox{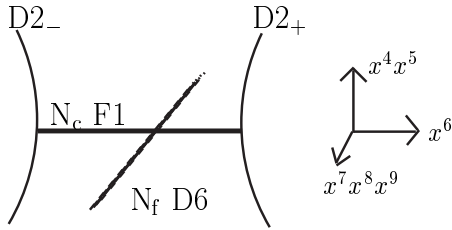}}
\caption{\small The combination
of (a) and (b) of Fig.\ref{D2pm}.}
\label{MQCD}
\end{figure}

Let us generalize the previous result to the case with two D2-branes and
$\Nc$ fundamental strings stretching between them. We consider the
configuration in which the $\Nf$ background D6-branes are
located between the two D2-branes. 
This is the similar situation to the MQCD configuration.
We can see that 
U(1)$_-$$\times$U(1)$_+$ gauge theory with heavy $\Nc$ bifundamental
quarks on the $\Nf$ D6 background is realized on this 
configuration.\footnote{
This is rough approximation and we know that
there is nonperturbative effect. But here, we proceed further keeping
in mind that this approximation is justified only
in the ultra-violet region $u \gg \Laq$ (QCD scale) 
or in the large $\Nc$ limit. We will go back to this problem later.}
Here, we distinguish each D2-brane and U(1) factor 
by $\pm$. 
This corresponds to the situation that the U(2) gauge theory is broken 
into the U(1)$_-\times$U(1)$_+$ gauge theory by the relative difference
between the nontrivial fields
${\tilde \ph_{\pm}}$ on the D2-branes.
Then, we treat the two D2-branes almost independently
except that in this situation, the signs of $\pm 2\Nc$ and 
sign$(\ph_0) \Nf$ appear in the combination as 
$(-2\Nc, \Nf)$ and $(+2\Nc, -\Nf)$. (see Fig.\ref{D2pm} and
Fig.\ref{MQCD}.)  

This leads to that the relative
distance 
$\De \{\ln y\}\equ (\ln y_+    - \ln y_-) $
is determined as
\eq
\De \{\ln y\} = \ln y_ +  -   \ln y_- 
= \fr{1}{2}\left\{\De {\rm \tilde \ph} + i \De \chi\right\}
=(2\Nc - \Nf) \ln \fr{u}{\La} + {\rm const.}\hsp{1cm}.
\en
This is the correct behavior for the RG-flow 
of SU($\Nc$) SQCD with $\Nf$ flavors. 
\section{T-dualized Configuration}
Until now, we have discussed the configuration which is analogous to
the MQCD configuration. 
We have studied the $2+1$ dimensional field theory realized
on this configuration.
In this section, we discuss the T-dualized
configuration in the direction of $x^6$. 
For our original purpose, we have to realize the equivalent
$2+1$ dimensional field theory also on the T-dualized configuration.
What does the previous D2-D6 system transform under the T-duality ?
Naively it seems to be D3-branes on the background of
the D7 SUGRA solution.
But, there might be some confusion
about what is the background after the T-duality.

In the previous attempts \cite{0106014}\cite{0107057},  
they have considered the Type IIB configuration with
the orbifold on the D7 SUGRA background.
They have regarded this configuration 
as the T-dual of the MQCD configuration with two 
NS5-branes on the D6 SUGRA background.
As a result of that, the logarithmic behavior of the
D7 SUGRA solution makes the analysis very messy. This
also hampers having the AdS$_5$ structure in the conformal case,
as found in \cite{0106014}. 
This logarithmic behavior is the origin of the abnormal (complicated)
behavior of their result. This
strongly suggests that the D7 SUGRA background will not be the  
correct background as the T-dual of the MQCD configuration.

The most important point is that the region where the D7 classical
solution is effective corresponds to that of the MQCD configuration with
the small $x^6$-radius. In this region, the two NS5-branes are wrapping
this direction and crossing each other. We can not expect that the
ordinary 4D gauge theory is realized on this configuration. So it is
unlikely that the role of the D6 SUGRA solution as the background will
simply be succeeded to the D7 SUGRA solution.

In other words, the background in Type IIB theory must have
the radius $R^{{\rm IIB}}$ which satisfies the relation
$R_6^{\rm IIA} = \Ap /  R_6^{\rm IIB} \ra \infty$.
This is the situation for the D6 background in the MQCD configuration.
In this sense, the D7 SUGRA solution is not equivalent
to the  D6 SUGRA background in the MQCD configuration. 
The D7-brane solution is obtained from the small $R^{{\rm IIA}}$
limit of the D6 solution.
On the other hand, this requirement is satisfied in the case of pure SYM theory;
the backgrounds are flat before and after the T-duality.   

Also in our simplified model (T-dual of the D2-D6 system), the D7 SUGRA
background will not be the correct background.  But it is very plausible
that the scalar field $X^6/2\pi\alpha^{\prime}$ on the D2-brane world
volume is transformed into the Wilson line (or gauge field) $A_6$ on the
D3-brane. Then, the form of the D3-brane action after such translation
enables us to guess at least how is the background which interacts with
the fields on the D3-brane.

Let us look back at the D2-brane action (\ref{D2action}) and consider
how the action will change after the plausible T-duality. 
After rewriting ($\ph$, $a$) by ($a_6$, $a_0$), where 
$a_6 \equ \fr{A_6}{\La}$ and $a_0 \equ \fr{A_0}{\La}$,
we expect that the D2-brane action reduces to that of the D3-brane with
the delocalized direction of $x^6$. The D3-brane action after the
dimensional reduction in this direction will be 
\eqn
S_{{\rm D3}} &=& - \fr{\La}{4\pi} \int dt du^4 du^5 {\rm H}  
 \Bigg[\fr{1}{\La^2} + \fr{1}{2} \Big|\vn_u a_6 \Big|^2 
    -  \fr{1}{2} \left|\vn_u a_0 \right|^2
                     + O (\La^{2})\Bigg] \nn
&& \hspace*{2cm} + \fr{\La^2}{4\pi}\int dt du^4 du^5 
          \fr{\Nf}{R^2} \left(\fr{a_6}{R} - 
               \vn_u a_6 \cdot \fr{\vec{u}}{R}\right)a_0 ,
\label{D3action}
\enn
where we define $R$ and H as 
$R \equ \sqrt{|u|^2 + \left(\La a_6\right)^2}$
and ${\rm H} \equiv 1 + \fr{\La \Nf}{R}$. 

Let us estimate  
the form of the 'background' on the D3-brane which 
gives the D3-brane action (\ref{D3action}).
It is known that the D3 brane action on the
general background can be written as,
\eqn
&&S_{{\rm D3}} = - T_{{\rm D3}} \int d^4 \sig \dilm 
  \sq{-{\rm det} \left(g_{M N } \pa_{\A} X^{M} \pa_{\B} X^{N} 
   + 2\pi \Ap F_{\A \B} \right)} + \fr{1}{4 \pi^2 \gs \Ap }
              \int G^{{\rm R}}_{(3)} \wedge A_{(1)} , \nn
&& \hspace*{2.5cm} G^{{\rm R}}_{(3)} \equ d B^{\rm R}_{(2)} 
       + \left(B^{{\rm NS}}_{(2)} 
                + {2 \pi \Ap F}_{(2)}\right)
                   \wedge d C^{\rm R}_{(0)},
\label{IIBaction}
\enn 
where we defined $T_{{\rm D3}}$ as the D3-brane tension 
$T_{{\rm {D3}}}= (8\pi^3 \gs {\Ap}^{2} )^{-1}$.
In the above equation, $B^{\rm R}_{(2)},B^{{\rm NS}}_{(2)} $ 
and $C^{\rm R}_{(0)}$ are the RR 2-form, NSNS 2-form and RR 0-form gauge
field respectively. 
First, we have to careful of the region where the field theory is the
good description.
Let us denote the radius in the direction of $x^6$
in Type IIB theory as $R_6$ (=$R_6^{{\rm IIB}}$) for simplicity. 
By using the relation of the string
coupling constant \footnote{We have used the same symbol $g_s$ to
express the string coupling constant for both IIA and IIB theory.  Here
we distinguish the two kinds of the string coupling constant.}  between
before (IIA) and after (IIB) the T-duality, $g^A_s =
{g^B_s}{{\Ap}^{1/2}}/{R_6}$, we can express $\La$ as $\La =
{g^A_s}/\left(4 \pi {\Ap}^{1/2}\right) = {g^B_s}/\left(4 \pi {R_6}\right)$.
Then in Type IIB theory, 
we can take the similar limit to (\ref{limit}) as
\eqn
 \gs \ra 0 \ \ , \hsp{0.6cm} 
R_6 \ra 0 \ \ , \hsp{1cm} &&  z \ra 0 \ \ , \hsp{0.6cm}\Ap \ra 0 \ \ ,\nn
\La  \equ \fr{\gs }{4 \pi R_6} \ \ ,  
\hsp{0.3cm}&& u \equ \fr{z}{2\pi \Ap} \hsp{0.5cm} {\rm fixed} \ \ .
\label{limitB}
\enn
In addition to the above limit, let us take the limit $\Ap / R_6 \ra \infty$.
This means that the compactified radius of $x^6$-direction becomes
infinite in Type IIA theory. This is the same situation as that in the
previous sections.

Then we can easily read off the 'background' from the actions
(\ref{D3action}) and (\ref{IIBaction}) in the limit of (\ref{limitB}).
The 'background' is written as
\eqn 
&&ds_{10}^2= {\rm H}^{-\fr{1}{2}}\left\{ \eta_{\mu \nu} dx^{\mu}dx^{\nu} 
+ \sum_{i = 7,8,9} (dx^i)^2 + (dx^6)^2 \right\}
+  {\rm H}^{\fr{1}{2}}|dz|^2  , \nn  
&& {\rm e }^{-\ph} =  
             {\rm g}_{\rm s}^{-1} {\rm H} ,
\label{BG} \\
&&\pa_4 B^{\rm R}_{56} - \pa_5 B^{\rm R}_{46}
                = \fr{\Nf a_6 \La^2}{2\pi \Ap R^3} , \hspace{1cm}
\vn C^{\rm R}_{(0)} = \fr{\Nf \La}{2\pi \Ap R^3}\rot{-u^5}{u^4}{} , 
\hspace{1cm}  B^{\rm NS}_{64} =B^{\rm NS}_{56} =0 ,
\no
\enn 
where we take the static gauge for the action (\ref{IIBaction}) as before.
Here we define $\vn$ as the derivative with respect to the original
coordinates of the 2D space $(x^4,x^5)$.  Note that in the above
expression, the above 'background' expresses the only gravitational field 
on the D3-brane. 
We can not guess how the background behaves in the
bulk away from the D3-brane. 
But we have to remember that only the geometry near the brane
is important for the AdS/CFT correspondence. So it is enough for
that purpose to obtain the information about the background on the brane.

Note that the inclusion of the Wilson line with periodicity means that
this expression contains all the winding modes of the
$x^6$-direction. 
This is equivalent to the fact that the D6-supergravity background contains
all the Kaluza-Klein (KK) modes in this direction. 
As the radius $R_6$ becomes large,
the effect of the non-zero winding modes drops and we have to change 
the warped factor as
\eq
\fr{\La \Nf}{R} \ra \sum_{n=-\infty}^{\infty} 
   \fr{\La \Nf}{\sqrt{|u|^2+ (\La a_6 + n/R_6 )^2}} 
\ \ \sta{|u| \gg 1/R_6}{\mbox{\LARGE{$\sim$}}} \ \ 
 \fr{\gs \Nf}{2\pi} \ln |u| + {\rm const.} \hsp{1cm}.
\en
Then the above 'background' becomes
the simple D7 supergravity solution with the non-trivial dilaton
and RR 0-form.
It is this simple D7 solution that has been used in 
\cite{0106014}\cite{0107057} as the background.
But in the region with the large radius $R_6$ (small radius $\Ap/R_6$ in Type IIA),
we can not expect any more that the result
of the MQCD analysis will be reproduced in Type IIB theory. 
This will be the reason why their result seems to be
different from the result expected from the 4D field theory. 
 
Here, we have to comment on the anomaly inflow mechanism \cite{9605033}
in the D3-D7 system 
with the two-dimensional intersection.
This means the cancellation between
the anomaly coming from the chiral fermion 
on the intersection and the anomaly from the bulk Chern-Simons term. 
In our model, the fermion one-loop effect is included
in the first term of Eq.(\ref{IIBaction}) and the bulk Chern-Simons
term corresponds to the second term. As for the first term, this
is the picture of the closed string.
These two contributions are canceled under the BPS condition
such as $\vn a_0 = - \vn a_6$ and the Gauss-Low. 

\begin{figure}
\epsfysize=6cm \centerline{\epsfbox{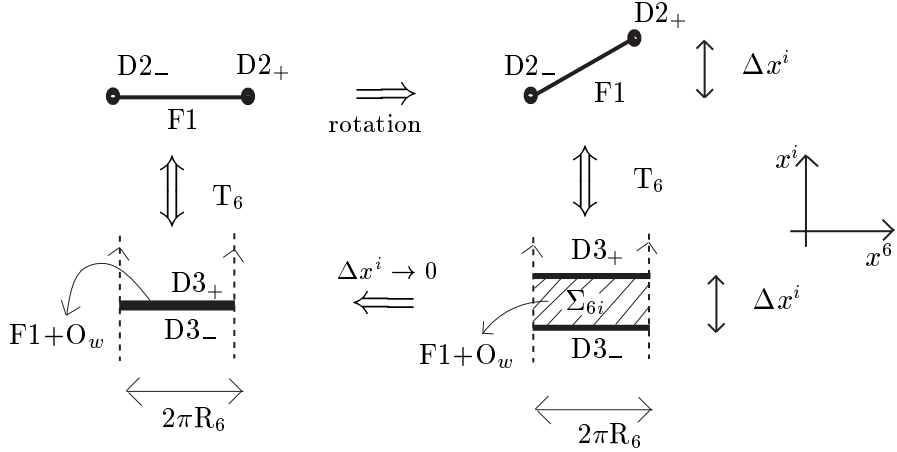}}
\caption{\small
The sketch in the case of $\Nf=0$. 
The number of the Type IIA fundamental strings transforms
to that of the Type IIB fundamental strings. They are stretching over the  
the vanishing distance expressed as the zero limit of the small 
resolution $\De x^i$.}
\label{resol}
\end{figure}

Next, let us discuss the behavior of the field on this D3-brane. 
We can repeat the same procedure as before by replacing
$(\ph, a)$ and $\ph_0$ in section 3
with $(a_6, a_0)$ and $\left.a_6\right|_{u=0} $.
In the same way, we define the field which expresses the F1 
density on the D3-brane as
\eq 
\tvn_u \chi \equ {\rm H}\vn_u a_0 + \Nf \left(
\fr{\La a_6}{R} -1\right)\fr{{\vec u}}{|u|^2} \hsp{3cm}
\tvn_u \equ  \rot{-\pa_5^u}{\pa_4^u}{}.
\en
What is the fundamental string like whose charge is described by the
above field $\chi$ ?
Let us consider the case with two D3-branes and $\pm \Nc$ additional
electric (F1) source on the flat background ($\Nf =0$).
This also expresses the U(1)$_-$$\times$U(1)$_+$ gauge theory with 
$\Nc$ bifundamental matters as discussed in the previous sections. 
The source is realized as the T-dual of the
fundamental string in the previous section. This fundamental
string in Type IIB theory is stretching over the vanishing distance between
the two D3-branes, say, the distance in the direction $x^i$ $(i=7,8,9)$.
As an example, the rough sketch is depicted as Fig.\ref{resol}. 
On the other hand, the information about the distance between the D2-brane
is transferred to the integral on the vanishing 2-cycle $\Sig(6i)$
of the NSNS 2-form field $\int_{\Sig(6i)}
B_{(6i)}^{{\rm NS}}$.
This integral is also 
the holographic charge\footnote{ 
Here 'holographic charge'
means the charge which is observed at $u$ in the same way
as the RG-flow of the gauge coupling constant in section 2.
This charge shows how the string winds around in the 
direction of $x^6$.}
of the wave O$_w$ 
per fundamental string mentioned in the above.

In the case with $\Nf \neq 0$, there are also $\Nc$
fundamental strings stretching over the vanishing distance. 
But different from the case with $\Nf = 0$, 
we have to generalize the NSNS 2-form field
in the same way as 
$\De {\tilde \ph}$ in the previous section.
This generalized  NSNS 2-form field gives the correct wave O$_w$
charge per fundamental string.
These topics will be discussed in the following sections. Here we limit
our analysis in this section within that of the world volume theory
and proceed further.

Assuming that the bifundamental matters are
too heavy to give dynamical effect, 
we can handle the two D3-branes almost
independently.
Then by using the similar complex coordinate $y$ to the previous sections,
\eq
\ln y \equ \fr{{\tilde a_6} + i\chi}{2} 
  =\fr{1}{2}\left\{a_6 +  i\chi + \Nf \ln 
   \left(\fr{R + \La a_6 }{\La}\right)\right\} ,
\label{a_6chi_sol} 
\en
we obtain the solutions and the difference between them as
\eqn
\ln y_{\pm} 
  &=& \pm \left\{\Nc - \fr{\Nf}{2}\right\} \ln \left(\fr{u}{\La}\right) 
+ \fr{\Nf}{2} \ln \left(\fr{u}{\La}\right) + {\rm const.}\nn
\De \{\ln y\} &=& \ln y_+ - \ln y_- =\fr{1}{2}\left(\De {\tilde a_6} 
+ i \De \chi\right)
=  \left(2\Nc  - \Nf\right)\ln  \left(\fr{u}{\La}\right)
+ {\rm const.}\hsp{0.5cm}  .
\label{a_6chi_solapdif}  
\enn

\section{From the Fields on the Brane to the Fields of Supergravity }
In the previous section, we have learned that the 
(generalized) rescaled Wilson line $a_6$ (${\tilde a_6}$) 
and the new field $\chi$ on the D3-brane give the non-trivial solution due to
the background in the action (\ref{D3action}). 
Let us consider rewriting 
these non-trivial fields 
in terms of the Type IIB SUGRA matter (gauge) fields.
This will be useful for the application to AdS/CFT
(gravity/field theory) correspondence. This is also the necessary procedure
because this teaches us how to transform these field under the 
sequence of T- and S-dualities.

What is the supergravity matter (gauge) fields 
corresponding to $a_6$ and $\chi$ ? 
First, let us consider the rescaled Wilson line $a_6$. 
It is well known that the
Wilson line on one D-brane can be measured by the string world sheet
coupled by NSNS 2-form field. This world sheet spans the circle of
the compactified direction ($x^6$ direction in our case) and the
orthogonal semi-infinite line from the point located by the D-brane. 
From the field theoretical point
of view, a string stretching on this
semi-infinite line expresses an external heavy quark on the D-brane.
Then we can obtain the Wilson line by the integral of the gauge 
field over the (compactified) circle, that
is, by the world sheet with NSNS 2-form field.

Let us discuss how to express the rescaled Wilson line $a_6$ 
in our model. 
First, let us consider $x^i$ as one of the coordinates $(x^7,x^8,x^9)$. They are 
in the orthogonal
directions to the D3-brane in the previous section.
We have the relation between the NSNS field and the Wilson line as
\eq
\fr{1}{2\pi \Ap}\int_0^{2\pi {\rm R_6}}  
  \int_{x^{i}=\infty}^{x^{i}=0}   dx^6dx^{i}    
B_{6i} = \oint_{x^{i} =0} dx^6 A_6 - 
      \oint_{x^{i} =\infty} dx^6 A_6 , 
\en
where we denote $x^{i}=0$ as the position
of the D3-brane in this direction. We set the Wilson 
line at infinity zero below.
Remember that we consider the location of the $\Nf$ D6-branes as the origin
in the Type IIA analysis. This means the vanishing 
Wilson line for the background.

Then we have a relation between $a_6$ on the D3-brane 
and the NSNS field as
\eq
\fr{\gs a_6}{2} = \oint_{x^{i}=0} dx^6 A_6 
= \fr{1}{2\pi \Ap}\int_0^{2\pi {\rm R_6}}  
  \int_{x^{i}=\infty}^{x^{i}=0}   dx^6dx^{i}  B_{6i}^{{\rm NS}} .
\label{adef}
\en
Let us define new fields, 
$\bns$ and ${\tilde \bns}$ as the NS fields
corresponding to $\left(\gs \De a_6\right)/{2} $
and $\left(\gs \De {\tilde a_6}\right)/{2}$ 
in the previous section.
They can be written as
\eqn
&& b^{{\rm NS}} \equ
\fr{1}{2\pi \Ap}\int_{\Sig (6i)} 
B_{6i}^{{\rm NS}}dx^6 dx^i 
   = \oint A_6^+ d x^6 - \oint A_6^- d x^6  
     = \fr{\gs}{2} \De a_6 , 
\nn
&&{\tilde b^{{\rm NS}} } \equ 
\fr{1}{{2\pi \Ap} }\int_{\Sig (6i)} 
{\tilde B_{6i}^{\rm NS}} dx^6 dx^{i}    
 \equ  \fr{\gs}{2} \De {\tilde a_6}
= b^{{\rm NS}} +
  \Nf \ln \left(\fr{{\rm R_+} 
    + \La a_{6+}}{{\rm R_-}
    +\La  a_{6-}} \right) \no
\enn
where 
$\fr{1}{2\pi \Ap}$ is the normalization
factor for the 2-cycle $\Sig (6i)$. As in the previous section, 
this 2-cycle spans 
the circle with the radius $R_6$ and 
the vanishing distance between the two D3-branes
in the $x^i$ direction (see Fig.\ref{resol}).
In the above, we distinguish the gauge field
$A_6$, $a_6$ and R on each D3-brane by giving $+$ or $-$ on it. 
Note that as seen from these equations, $B_{6i}^{{\rm NS}}$ and 
${\tilde B_{6i}}^{{\rm NS}}$ 
have only $x^i$
dependence as the delta function and has nontrivial
$(x^4,x^5) $ dependence.

Next, let us discuss what is the supergravity fields corresponding to
$\De \chi$. Remember that $\chi$ measures the difference of 
the string density on the D3-brane. 
As discussed in section 4, there are $\Nc$ 
fundamental strings stretching over the 
vanishing distance between the two D3-branes in the direction
of $x^i$ ($i=7,8,9$). Then we can see that the NSNS 2-form 
related to the F1 charge will be the 
correspondent to $\De \chi$. 
From the equation which gives the F1
charge on $(x^4,x^5)$ space, we get   
\eqn
&&\hspace*{-1cm}\fr{1}{4\pi} \pa_\A \left(\De \chi\right) 
               = \fr{1}{(4\pi^2 \Ap )^3} \int
        \left(*H_{(3)}^{{\rm NS}}\Exp^{-\ph}\right)_{\A 1236jk}
        dx^{123} dx^6 dx^{jk}
    = \fr{1}{(4\pi^2 \Ap )^3} \int
    H^{{\rm NS}}_{\A 1236jk}
        dx^{123} dx^6 dx^{jk} ,\no
\enn
where the indices $\{j,k\}$ are in $\{7,8,9\}$, but $\{j,k\}\neq i$, 
and the index $\A$ is in $\{4, 5\}$. In the above, 
$\fr{1}{({4\pi^2 \Ap})^3}$ is the 
normalization to give the integer F1 charge.
We also use the last expression as the Poincare 
dual of the NS-NS 3-form field strength. 
In the T-dualized (Type IIB) model, 
the directions of $\{x^{0123}, x^6\}$ are 
delocalized.\footnote{As we will
see, we will take the T-duality in these directions.} 
So we can see that this 7-form $H^{{\rm NS}}_{\A 1236jk}$ has  
$(x^j,x^k)$ dependence as the delta function in addition to nontrivial
$(x^4,x^5) $ dependence.

Then we can write down the solution in terms of the fields 
of supergravity as 
\eqn
&& 
\fr{1}{2}\left\{\De {\tilde a_6} + i \De \chi\right\}  =
   \fr{1}{2\pi \Ap}\int_{\Sig (6i)} 
     \gs^{-1} {\tilde B_{6i}}^{{\rm NS}}dx^6 dx^i 
       + \fr{i}{(2\pi)^5 {\Ap}^3} \int
          B^{{\rm NS}}_{ 1236jk}
              dx^{123} dx^6 dx^{jk} \nn 
&& \hspace*{2cm}
=  \left(2\Nc  - \Nf\right)\ln u + {\rm const.} \hsp{0.5cm},
\label{bns_sol}
\enn
where the NSNS 6-form gauge field $B^{{\rm NS}}_{ 1236jk}$ 
is defined as $\pa_{\A} B^{{\rm NS}}_{ 1236jk} =  
H^{{\rm NS}}_{\A 1236jk}$.
Note that the above supergravity gauge fields are living only  
'between' the two overlapping D3-branes and similar
to those of the twisted sector on the orbifold.
But we have to be careful of the fact that
the above real part is not written only by 
$\fr{1}{2\pi \Ap}\int_{\Sig (6i)} 
     \Exp^{-\ph}B_{6i}^{{\rm NS}}$, which is 
different from the case of pure SYM theory.
Note that the integral $\fr{1}{2\pi \Ap}\int_{\Sig (6i)} 
     \gs^{-1} {\tilde B_{6i}}^{{\rm NS}}dx^6 dx^i$ is also
the holographic wave charge per fundamental
string.

It will be interesting to bring our results to the configuration
in which the four dimensional gauge theory is realized. 
This is the topic of the next section.

\section{4D ${\cal N}$=2 Field Theory and Gravity Solution}
\subsection{Gauge Coupling Constant and 2-Form Fields}
Let us take T-dualities and S-dualities of our configuration,
and make the model in which 4D ${\cal N}$=2 SQCD is realized.
This is the T-dualized model obtained from the well-known
MQCD configuration mentioned in section 2 and our result 
will give us some knowledge of what it is like. 

We consider the sequences of the dualities, ${\rm
T_{36}ST_{12}ST_{36}}$, where the indices mean the directions of which
we take T-dualities.
Remember that the directions $\{x_1,x_2, x_3\}$ do not play any active role in our
analysis. On the D2 world volume, the three
scalar fields for these directions 
are free and decoupled from the remaining interacting action
(\ref{D2action}). In fact, we can easily
confirm that there is no warped
factor H on their kinetic terms.
So we can safely delocalize 
these directions without changing
our analysis.
This is also the reason why the two-dimensional supersymmetric cycle
for the M2-brane on the Taub-NUT background
is the same as that of the M5-brane on the
same background.
As for the direction of $x_6$, we have already 
discussed the T-duality in this direction with special care in the
section 4.
The other directions $\{x_4,x_5, x_7, x_8,x_9\}$ 
are important for the structure of the 
vacua of the 4D field theory, but we do not take the T-dualities
of these directions.

Let us consider what the constituents in our model
will transform into. 
They are expected to transform under these dualities as
\eqn
   \fr{1}{2\pi \Ap}\int_{\Sig (6i)} 
     \gs^{-1} {\tilde B_{6i}^{{\rm NS}}}dx^6 dx^i  \hsp{1cm} 
 &\ra& \hsp{1cm}
   \fr{1}{2\pi \Ap}\int_{\Sig (6i)} 
       \gs^{-1} {\tilde B_{6i}^{{\rm NS}}}dx^6 dx^i \nn
 \hsp{1cm} \fr{1}{(2\pi)^5 {\Ap}^3} \int_{\Sig (jk)}
          B^{{\rm NS}}_{ 1236jk}
              dx^{123} dx^6 dx^{jk}
\hsp{1cm}&\ra& \hsp{1cm}
   \fr{1}{2\pi \Ap} \int_{\Sig (jk)}
          B^{{\rm R}}_{ jk} dx^{jk} \nn
 \hsp{1cm}{\rm  D3}(456) \hsp{1cm}&\ra& \hsp{1cm}
{\rm  Kaluza-Klein\ \ monopole}(12345) \nn
 \hsp{1cm}{\rm N_c\ \ F1}(i) \hsp{1cm}&\ra& \hsp{1cm}
{\rm N_c\ \ D5}(1236i) \nn
\hsp{1cm}{\rm  O}_w(6) \hsp{1cm}&\ra& \hsp{1cm}
{\rm  D3}(123). \label{tdual}
\enn
In the above, we can see how the new NSNS 2-form
${\tilde B_{6i}^{{\rm NS}}}$ transforms
by Eq.(\ref{adef}) and the property 
$B^{{\rm NS}}_{6i} \ra B^{{\rm NS}}_{6i}$ under 
this transformation.
We also have to mention that the background does not
change as seen from the explicit form (\ref{BG}).
Note that the $\Nc$ strings stretching over the
vanishing distance between the two D3-branes 
transform into $\Nc$ D5-branes (1236i), which are 
also wrapping the vanishing 2-cycle(6i) between the two 
Kaluza-Klein (KK) monopoles. In addition to that, because of the
existence of NSNS 2-form ${\tilde B_{6i}^{{\rm NS}}}$, there is also
induced D3-brane$(123)$ charge in the D5-brane in the same
way as the wave in the previous section. Due to this induced
charge, there is non-trivial RR 5-form flux. 
We also remark that the 4D gauge coupling constant corresponds
to the field $\int_{\Sig (6i)} \gs^{-1} {\tilde B_{6i}}^{{\rm NS}}dx^6 dx^i$
and that this is not written only by $\int_{\Sig (6i)} \Exp^{-\ph} 
B_{6i}^{{\rm NS}}dx^6 dx^i$. By Eq.(\ref{bns_sol}), 
we obtain the solutions for the above NSNS and RR 2-forms 
in the complex form as   
\eq
-i\gamma \equiv   \fr{1}{2\pi \Ap}\int_{\Sig (6i)} 
       \gs^{-1} {\tilde B_{6i}^{{\rm NS}}}dx^6 dx^i + 
   \fr{i}{2\pi \Ap} \int_{\Sig (jk)}
          B^{{\rm R}}_{ jk} dx^{jk}
=  \left(2\Nc  - \Nf\right)\ln u + {\rm const.} \ \ \ .
\label{solution}
\en
This is the modified twisted sector of the 2-forms on the
background. 

\subsection{Gravity Dual: Suggestion}
Let us discuss the supergravity dual corresponding 
to this configuration. 
For this purpose, let us reconsider the MQCD
configuration first. It consists of
$\Nf$ rigid D6-branes, and
(NS5,D4)-branes.
The state or shape of (NS5,D4)-branes is
determined by the
BPS condition on the $\Nf$ D6 supergravity background.
The important point is that the shape of these 
(NS5,D4)-branes has the information about the field theory
dynamics.
Therefore, in order to discuss the gravity dual, 
we have to extract or separate the gravity induced
by these (NS5,D4)-branes from the background.
This is very difficult task.
But remember that the solution of the $\Nf$ D6 SUGRA solution
becomes $\Nf$ KK monopole solution in the eleven-dimensional supergravity.
Moreover, after the large $\Nf$ limit, this reduces to ${\rm Z}_{\Nf}$
orbifold, that is, locally flat metric \cite{9802042}\cite{9812159}. 
This simplifies the problem, and 
it will be possible to carry out the above extraction.
In general, the locally flat metric transforms another
locally flat metric under the T-duality.
So the above observation indicates that also in Type IIB configuration,
there is such a frame in which the background becomes locally flat. 

In addition to that, we have to remember that it is only the relative
(generalized) distance $\De {\tilde x^6}$ 
between the two NS5-branes that has the physical
meaning as the RG flow of the (complex) gauge coupling 
constant.
The position itself in the direction of ${\tilde x^6}$
does not effect on the 4D field theory.\footnote{
This direction has the physical meaning
for the 4+1 dimensional field theory on the D4-branes.
For the line-compactified theory (3+1 dimensional theory), this direction
loses the physical significance except the relative compactified length.
In fact, the MQCD supersymmetric cycle is determined up to the scale and 
phase transformation of the holomorphic coordinate $y$.
This transformation changes the form of the Seiberg-Witten curve, but 
does not change the mass formula for the soliton.
These degrees of freedom originates from the ones that
we can choose the origin anywhere for $(x^6,x^{10})$ space.}
This means that there is one extra degree of freedom
for the 4D field theory.
This enables us to delocalize the configuration 
in this direction, keeping the relative distance fixed.
This also simplified the problem.

As a conclusion, we can say that the problem will become easy in the 
following procedure:\\
(1) By adding the extra dimension, set the background to be the locally
flat.\\
(2) On this background, delocalize the configuration in the
irrelevant direction for the 4D \hspace*{0.6cm}field theory. 
\vsp{0.6cm}

But in Type IIB theory, the reliable higher dimensional effective theory
is not known. This
is the different point from Type
IIA theory related to the eleven-dimensional supergravity.  
So our following analysis is based on only the analogy of Type IIA
theory and the result is limited within the suggestion of the procedure 
to obtain the possible dual of the corresponding 
field theory. 

Let us return to our model in Type IIB theory and consider the problem
in the same spirit as the above.
What is the appropriate parameter which should be promoted to
the additional space coordinate ? 
The electric field (or the temporal component of the gauge field)
on the D-brane
will be the promising candidate. This is because this field is known to
have the relation with the eleventh dimension 
in Type IIA, as discussed
in the previous sections. 

On the other hand, the coordinate $x^6$ in Type IIB theory 
does not play any active role.
The configuration is delocalized in this direction and the role of the
coordinate $x^6$ in the MQCD analysis is succeeded to the Wilson line.
As a result of that, Type IIB theory we have been discussed is the
almost nine-dimensional theory. So let us promote also the Wilson line
to the new coordinate.  Then this almost nine-dimensional theory is on
the same level with the ten-dimensional Type IIA theory on the point of
the degree of freedom for the space-time dimension.

We have to note that the two-dimensional space ($x^6,x^{10}$)
in M-theory is related by the T-duality to the torus with the complex structure
$\tau \equ C_0 + i/\gs$ in Type IIB theory \cite{9508143}.
So we can expect that the electric field and
the Wilson line play the role of the new coordinates of this additional
two-dimensional space for Type IIB theory.

On the KK monopole, the Wilson line and the electric field correspond 
to the NSNS and RR 2-forms respectively as already seen in the sequence
of the T- and S-dualities. 
These two parameters can be observed only on the branes
in Type IIB theory.
But, by including them as the new space coordinates,
we can formally
extend our discussion to twelve-dimensional space-time.
Let us define the new coordinate as
\vspace{-0.3cm}

\eq
x^{{\rm NS}} \equiv \fr{a_6 \gs }{2},\hsp{2cm}
x^{{\rm R}} \equiv \fr{\chi}{2}.
\label{newco}
\en
Note that $x^{{\rm R}}$ and $x^{{\rm NS}}$ are the periodic 
coordinates. 
But in our discussion, the Wilson line $x^{{\rm NS}}$ always
appears in the form $d{\tilde x_{{\rm NS}}}\equ \Exp^{-\ph} dx_{{\rm NS}}$
with the string coupling constant $g_s
\ra 0$. So the coordinate ${\tilde x_{{\rm NS}}}$ runs from $-\infty $ to
$+\infty$ in the same way as $a_6$ or ${\tilde a_6}$.
This also means that 
there is no S-invariance of the SL(2 Z) and S-transformation is fixed
\footnote{
This is also seen from the fact that the radius 
in $x^6$ direction is infinite in Type
IIA MQCD configuration, there is no symmetry to exchange the radiuses
for the directions of $x^6$ and $x^{10}$.} 
in our analysis. 

How can we lift the ten-dimensional supergravity solution to
the twelve-dimensional solutions ?
The hint is given by
the T-invariance of the SL(2 Z)
and the analogy of the lift
from Type IIA to eleven-dimensional supergravity.
We suggest the form as \footnote{
The factors $\gs^{1/2}$ and $\gs^{-1}$ are required 
for our convention in order to kill $\gs$ dependence
in $\Exp^{-\ph/2}$ and $\Exp^{\ph}$ respectively.} 
\eqn
ds_{12}^2 &=& \Exp^{-\ph/2 } \gs^{1/2}ds_{10}^2 
           + \Exp^{\ph}\gs^{-1}ds_2^2 \label{lift} \\
&& ds_2^2 \equiv \left(\fr{\gs\Ap}{R_6}\right)^2
\left[\Exp^{-2\ph} dx_{{\rm NS}}^2 + \left\{dx_R - dx^M\oint\left(
 B^{{\rm R}}_{6M} - C_{0}B^{{\rm NS}}_{6M}\right)
     \fr{dx^6}{2\pi \Ap} \right\}^2\right]
\no
\enn
where $M$ means the indices for the ten dimensional space-time
and runs from $0$ to $9$.  
The factor $\fr{\gs\Ap}{R_6}$ corresponds to the radius
of the eleventh dimension in Type IIA theory. 
We can see that by the above warped factor of the dilaton,
the Einstein action $\int \sqrt{-g^{(12)}} R^{(12)}$ in the twelve 
dimensions reduces to the
ten-dimensional action in the string frame
$\int \sqrt{-g^{(10)}} \Exp^{-2\ph}R^{(10)}$.

Note that the degrees of freedom for the metric are
the same as eleven-dimensional supergravity. This is because the 
relative factor for $dx_{{\rm NS}}^2$ and $dx_{{\rm R}}^2$
is fixed by T-invariance and there is the constraint 
that all the fields (and metric) are independent
of $x^6$ with the isometry. The latter reason comes from the 
requirement of the T-dual of Type IIA
theory. 

This is the important point.
There is the well-known fact that there is no supergravity
theory in the full twelve-dimensional space-time. But we have to note
that the above expression is defined only under the above conditions. 
As a result of that, the above 
expression is essentially eleven-dimensional one.
So the no-go theorem in the full twelve-dimensional space-time
does not mean that supersymmetry does not exist in our model. 

By this lift rule, the 'background'   
Eq.(\ref{BG}) becomes the KK-monopole solution,
\eqn
&&\hspace*{-2cm}ds_{12}^2 = \eta_{\mu\nu}dx^{\mu}dx^{\nu} 
+ \sum_{i = 7,8,9} (dx^i)^2 + (dx^6)^2 
+  ds_4^2 ,\label{KK} \\
&& \hspace*{-1cm} ds_4^2 \equiv \left(2\pi \Ap\right)^2 
\left\{
  {\rm H} \left(|du|^2 + \La^2 da_6^2  \right) 
     + {\rm H}^{-1} {\La^2} \Big[d\chi 
      + \Nf \left(\fr{\La a_6}{R} -1\right)d\varphi\Big]^2
\right\},
\no
\enn
where we use the same notation used in Eq.(\ref{chi}) and Eq.(\ref{BG}).
Then, let us take the large
$\Nf$ limit along with the limit in the previous sections.
By this limit, the warped factor H reduces as ${\rm H}=1 + \La \Nf / R \ra 
\La \Nf / R$. Then 
we obtain the locally flat metric as
\eqn
&&ds_{12}^2 = \eta_{\mu\nu}dx^{\mu}dx^{\nu} 
+ \sum_{i = 7,8,9} (dx^i)^2 + (dx^6)^2 
+  \left(2\pi \Ap\right)^2 \left|d{\cal M}^{(4)}_{{\rm Z_{\Nf}}}\right|^2  , \nn
&&\hspace*{1.5cm}   \left|d{\cal M}^{(4)}_{{\rm Z_{\Nf}}}\right|^2
    \equiv \left|dV_1\right|^2 + \left|dV_2\right|^2 , \nn
&&\hspace*{-0.5cm} V_1 \equiv 
    \left(2\Nf\right)^{1/2} \La \left(a_6 + \fr{R}{\La}\right)^{1/2} 
        \Exp^{i\chi/2\Nf}, \hspace{0.3cm}
V_2 \equiv  
     \left(2\Nf\right)^{1/2} u \left(a_6 + \fr{R}{\La}\right)^{-1/2} 
      \Exp^{-i\chi/2\Nf}.
\label{V12}
\enn
Note that in our notation, $\chi$ has the period $4\pi$ and this leads
to the $Z_{\Nf}$ orbifold identification
$(V_1,V_2) \sim (\Exp^{2\pi i/\Nf}V_1,\Exp^{-2\pi i/\Nf}V_2)$.
We can see the above complex coordinates $\{V_1,V_2\}$ have relations with 
the holomorphic coordinates $\{y,w\}$ of the Taub-NUT space as
\eqn
&&\hspace*{-1cm} y \equiv 
    {\rm e}^{\left(a+ i \chi\right)/2}  \left(a_6 + \fr{R}{\La}\right)^{\Nf/2} 
       \hspace{2.5cm}\stackrel{\Nf \ra \infty}{\sim} \ \ \ 
\left(2\Nf\right)^{-\Nf/2} \left(\fr{V_1}{\La}\right)^{\Nf} ,\nn
&&\hspace*{-1cm} w \equiv 
    {\rm e}^{-\left(a+ i \chi\right)/2} \left(\fr{u}{\La}\right)^{\Nf} 
       \left(a_6 + \fr{R}{\La}\right)^{-\Nf/2} 
       \ \ \ \ \ \stackrel{\Nf \ra \infty}{\sim} \ \ \ 
\left(2\Nf\right)^{-\Nf/2} \left(\fr{V_2}{\La}\right)^{\Nf}\no . 
\label{yw}
\enn
Then, our problem reduces to the embedding of the 
three kinds of 'matter' although they are originally undivided.
\vsp{0.5cm}

(1)the two KK monopoles 

(2)the complex 2-form Eq.(\ref{solution}) which 
exists between them 

(3)the D3-brane charge induced in the $\Nc$ external D5-branes
\vsp{0.5cm}

Let us consider the contribution coming from each part
and discuss how to construct the gravity dual.
First, let us concentrate on the two KK monopoles.  
Note that in section 4, we can see that the sum for ${\tilde a_6}+ i\chi $
on each D3-brane is also non-trivial. We can see from 
Eq.(\ref{a_6chi_sol}) and Eq.(\ref{a_6chi_solapdif}), 
\eq
 \ln y_{{\rm S}} \equiv \ln  y_+ + \ln y_- 
= \fr{1}{2}
   \left\{{\tilde a_{6+}} + {\tilde a_{6-}}  + 
i \left( \chi_+ + \chi_-\right)\right\}
= \Nf \ln \left(\fr{u}{\La}\right) + {\rm const.}\hsp{2cm}
\label{sum}
\en
After the large $\Nf$ limit, we obtain the 
corresponding position of the whole of the two D3-branes as
$\Nf\ln V_2^{{\rm S}}$= constant.  
In the same way, after the sequence of the
T- and S-dualities in the section 6.1, we can reach the same conclusion$-$ 
The whole of two KK monopoles are located 
at $\Nf\ln V_2=\Nf\ln V_2^{{\rm S}}$
: constant. Of course, there remains the relative distance 
which corresponds to the 
complex 2-form Eq.(\ref{solution}). Let us leave the contribution
from this 2-form for the next discussion and concentrate on the
contribution from the two KK monopoles themselves. 

Note 
that we can not distinguish
the direction of the KK monopole world volume from the direction
in which the KK monopole charge is delocalized or distributed. 
For example, there are two kinds of the Type IIA KK monopole 
from the point of M-theory. One has the world volume in the direction
of the eleventh dimension and the other is delocalized in this direction.
The former type is obtained 
by the dimensional reduction from the KK monopole in M-theory 
with respect to the eleventh dimension. We can obtain the latter type by
taking the T-dualities from the Type IIA NS5-brane, for example,
T$_{56}$ dualities from the Type IIA NS5-brane(12345). But, the both types of
the Type IIA KK monopole are described by the same classical solution in
M-theory.

This means that when we delocalize the whole of the two KK monopoles 
in the direction of $\Nf \ln V_2$, we can obtain 
the ordinary 
Type IIB KK monopole solution which is non-trivial only in the directions 
$\{x^6,x^7,x^8,x^9\}$ and has the KK monopole charge with respect to
the compactified direction of $x^6$.
In the region where $x^i \sim 0$ ($i=7,8,9$), it is known 
that the supergravity solution
for the two overlapping KK monopoles reduce to the orbifold 
${\rm R^4/Z_2}$ \cite{9802042}.\footnote{
This is the same procedure as we have done for the $\Nf$ KK monopole
solution.}

Note that on the space of $V_2 = V_2^{{\rm S}}$: constant, 
$V_1$ is the same as $u$
from the definition (\ref{V12}). 
So we can interpret $u$ in (\ref{solution}) as $V_1$ on the plane, 
$V_2 =V_2^{{\rm S}}$: constant.

Therefore, our problem reduces to the embedding of the remaining
two kinds of 'matter' into the locally flat background such
as
\eq
ds_{12}^2 = \left\{\eta_{\mu\nu}dx^{\mu}dx^{\nu} 
+ \left(2\pi \Ap\right)^2\left|d{\cal M}^{(4)}_{{\rm Z_2}}\right|^2
+ \left(2\pi \Ap\right)^2\left|d V_1\right|^2\right\} + 
\left(2\pi \Ap\right)^2\left|d V_2\right|^2
\label{orb}
\en
where we use $\left|d{\cal M}^{(4)}_{{\rm Z_2}}\right|^2$ 
as the symbol which expresses 
the 4D locally flat space with the $Z_2$ orbifold 
identification.
The two kinds of 'matter' are given as
\vsp{0.5cm}

(1)the twisted sector on the ${\rm R^4 /Z_2}$ orbifold fixed point
\eq
-i\gamma = \Nf\left(\ln V_2^- - \ln V_2^+\right)
       = (2\Nc -\Nf) \ln V_1 + {\rm const.}
\en
\hspace*{0.5cm}
(2)the D3-brane charge induced in the $\Nc$ external D5-branes wrapping the
vanishing \\
\hspace*{1.2cm}two cycle on the orbifold
\vsp{0.5cm}

Note that all the fields becomes independent of $V_2$ 
after being delocalized in this direction. 
As a result of that, this extra two-dimensional space
does not play an important role for the remaining 
ten-dimensional theory.  
So we can conclude that what we have done is to add 
the extra two-dimensional space (\ref{newco}) and to pick up the 
unimportant another two-dimensional space ($V_2$-space) from
the twelve dimensional space-time. 
This is the procedure
similar to the M-theory flip.
In this remaining ten-dimensional space-time,
the generalized twisted sector becomes the ordinary one. 

On the other hand, in F-theory it is known that the
extra two-dimensional space corresponds to the space
for the dilaton and axion of Type IIB theory. 
In the context of F-theory, our procedure is
the replacement of the two-dimensional space for the non-trivial 
dilaton and axion with another
two-dimensional space for the constant dilaton and axion.
That is, we take the frame of the (remaining) ten-dimensional 
in which the dilaton and axion are constant.
Our suggestion is that this remaining 
ten-dimensional space-time would be the dual of the corresponding
field theory. 

We also have to comment on the fundamental region of the
4D $Z_{\Nf}$ orbifolded space. The fundamental region  
can be taken as ${\bf C} \times {\bf C/Z}_{\Nf}$. When we
delocalize the configuration in the $V_2$-space, the region
of this space is ${\bf C/Z}_{\Nf}$ because it is in the form of
$\Nf \ln V_2$ coming from Eq.(\ref{sum}) that we delocalize  
the configuration.\footnote{We have to 
emphasize that it is keeping $V_1$ fixed when
we delocalize the configuration in the $V_2$ space.
This requires another (discreet) phase transformation
for $V_1$, according to the $\Nf$ regions of the $V_2$ space.
This kills the phase transformation of $V_1$ by 
the original ${\bf R^4/Z}_{\Nf}$ identification.}
As a result of that, the $V_1$-space spans
the whole complex plane. In other words, the $Z_{\Nf}$ orbifold
identification is invisible for the remaining ten-dimensional 
space-time.

Then we can see that the above configuration is the same as that of
pure SYM theory except the values of the D5 and
D3-charge. This is consistent with the fact that at one-loop level, the
structure of pure SYM vacua is qualitatively the same as the Coulomb
branch of SQCD. 

The ten-dimensional solution can be obtained by modifying the result 
for pure SYM \cite{0011077}.
They have discussed the supergravity solution for the
N D5-branes wrapping on the vanishing two-cycle on the fixed point 
of the ${\rm R^4/Z_2}$ orbifold.
But with only a bit of change about the D-brane charge, 
we can formally generalize their 
result.

The result is summarized as follows:
\eqn
&& \hspace*{0.5cm} d s^2_{12} = ds^2_{10} 
      + \left(2\pi \Ap\right)^2\left|d V_2\right|^2
\label{grasol}\\
&& \hspace*{1cm}\left({2\pi \Ap}\right)^{-1} d s^2_{10}  
  = \fr{\rho^2}{f(|V_1|,v)^{1/2}} 
      \left\{\eta_{\mu \nu} dx^{\mu} dx^{\nu}\right\}
          + \fr{f(|V_1|,v)^{1/2}}{\rho^2} \left\{|dV_1|^2 + 
             d {\cal M}_{{\rm Z_2}}^2 \right\} , \nn
&& {f(|V_1|,v)} = 
    8\pi \gs {\rm Q_{D3}} + 
     2\left(2\pi \gs  {\rm Q_{D5}}\right)^2
      \left[\ln \left\{\fr{\rho^4}{v^2 } \right\}
          + \fr{|V_1|^2}{v^2} + {\rm const.} \right] ,
\nn
&& \hspace{4cm} \rho^2 \equ |V_1|^2 + v^2, 
\hspace{1cm} v \equ \fr{R_6^{1/2}}{\Ap \pi} 
\left(\sum_{i=7,8,9}{x^{i}}^2\right)^{1/4},\nn
&& \hspace*{0.5cm} C_{(4)} = \rho^4 {f(|V_1|,v)}^{-1}  
    dx^{0} \wedge dx^{1}
          \wedge   dx^{2} \wedge dx^{3} ,
\hspace{1cm} B^{{\rm R}}_{(2)} = b^{{\rm R}} \omega_{(2)}, 
\hsp{0.5cm}  {\tilde B^{{\rm NS}}_{(2)}} 
       = {\tilde b^{{\rm NS}}} \omega_{(2)}, \nn
&& \hspace*{2cm}
   \gamma \equ \tau {\tilde b^{{\rm NS}}} -  b^{{\rm R}} =  i Q_{{\rm D5}}
        \ln V_1 + {\rm const.} 
\hspace{1cm} \tau = \fr{i}{\gs} + C_0 \ \ {\rm :} \ \ \ 
{\rm const.}  \hsp{1cm} , \no 
\enn
where we denote D3 and D5 charges as ${\rm Q_{D3}}$ and $Q_{{\rm D5}}$.
In their case of pure SU($\Nc$) SYM, 
the D5 charge is $Q_{{\rm D5}}=2\Nc$, and for the D3 
charge they have suggested $Q_{{\rm D3}}=\Nc /2$.
We have replaced the original coordinate $u$ with $V_1$ as explained.
In the above equation, we denote the 2-form 
which is dual to the vanishing 2-cycle $\Sig$ as $\omega_{(2)}$.
We normalize this
integral of the 2-form over $\Sig$, as $\fr{1}{2\pi \Ap} \int_{\Sig}
\omega_{(2)}=1$. This 2-form also satisfies the anti-selfduality condition,
$\omega_{(2)} = -*\omega_{(2)}$. The components of 
the above NSNS and RR 2-forms 
$\big\{{\tilde B_{(2)}^{\rm NS}}, B_{(2)}^{\rm R}\big\} $
are essentially the same as 
${\tilde B_{6i}^{\rm NS}}$ and $B_{jk}^{\rm R}$,
that we have obtained by T- and S-dualities in (\ref{tdual}).
In the following discussion,
we set the RR 0-form $C_0$ to be zero. 

Then let us consider the supergravity solution for our configuration.
It is easy to see that in our case, the D5 charge 
is $Q_{{\rm D5}} = 2\Nc - \Nf$.
What about $Q_{{\rm D3}}$ ? 
As we have commented before, this charge is determined by the
NSNS 2-form field ${\tilde B_{6i}^{\rm NS}}$ 
on the $\Nc$ D5-branes. We approximate
this as
\eq
Q_{{\rm D3}}=\fr{\Nc }{4\pi^2 \Ap}\int_{\Sigma}
\left.{\tilde B^{{\rm NS}}_{(2)}}\right|_{V_1=\ep} = 
  \fr{\Nc}{2\pi} \left.{\tilde b^{{\rm NS}}}\right|_{V_1=\ep} ,
\en
where $\ep$ means the low energy cut-off of $V_1$  
in order to avoid the region where 
${\tilde b^{{\rm NS}}}$ vanishes.
Note that our approximations in section 3 and 4
about the bifundamental matters are 
broken down in this region. 
This is because they are not heavy any more and become 
massless. This is the typical limit for the
perturbative analysis.  
The nonperturbative effect will cure
this kind of singularity.
Then we can also expect that 
the above D3-charge is determined by the low energy
effective coupling
constants for the U(1)$^{\Nc -1}$ gauge theories 
coming from the broken SU($\Nc$) gauge symmetry. 
We need the Seiberg-Witten curve to determine these coupling constants, but
this is beyond our current analysis.

We also have to comment on the scale of the Higgs branch. 
In the directions of $\{x^7,x^8,x^9\}$, we need the same limit
as that of the MQCD analysis such as 
\eq
x^i \ra 0 \hsp{3cm} {\tilde v^2} \equ 
  \fr{1}{{\Ap}^{3/2}} \left(\sum_{i=7,8,9} {x^i}^2\right)^{1/2}
\ \ \ {\rm fixed}.
\label{Higgs}
\en  
This scale ${\tilde v}$ corresponds to the directions of 
the vacuum expectation
values for the quarks in the fundamental representations, and does
not depend on the string coupling constant. In fact, the Higgs branch 
is known to have no quantum correction.
Compared with $v$, we can easily see that ${\tilde v}\gg v \sim 0$ in our
model (see (\ref{limitB}) and (\ref{Higgs})). 
Then only $V_1$ dependence remains in Eq.(\ref{grasol}) because
${\tilde v}$ dependence never appears in this solution.
Especially $\rho$ is determined by only $V_1$ and the holographic
energy scale is expected to be $V_1$.
These facts show that the above result corresponds to the Coulomb branch,
not Higgs branch. 

Note that the complex field 
$\gamma \equ \tau {\tilde b^{{\rm NS}}} -  b^{{\rm R}}$
corresponds to the complex gauge coupling
constant of the gauge theory. This theory is realized on the $\Nc$ D5-branes wrapping
on the vanishing 2-cycle. 
In our analysis, the NSNS 2-form is generalized as compared with
the ordinary one on the flat background (pure SYM), but reproduces the 
correct behavior of the gauge coupling constant for SQCD.

In \cite{0106014} \cite{0107057}, it is suggested that this typical ratio 1/2 
between $\Nc$ and $\Nf$ originates from the constant  
$b^{{\rm NS}}/2\pi=1/2$ on the orbifold.\footnote{
This is based on the result that the constant $b^{{\rm NS}}/2\pi$
on the orbifold is obtained as 1/2 when the perturbative 
string sigma model is used \cite{9611137}. 
But as discussed in \cite{0104026}, in general, we can have
an arbitrary value in the region $2\pi > b^{{\rm NS}} \geq 0$. 
This is also seen in the fact that this parameter corresponds
to the arbitrary distance between the two NS5 branes in Type IIA theory.}
This is an interesting suggestion, but it seems to be different from
our result about the RG flow. Our result is independent of this value.
Moreover, in their model, this typical value of the 
NSNS 2-form field induces the D5-brane charge in the 
world volume of the $\Nf$ D7-branes.
In our model, the induced D5 charge is expected to come 
from the two Kaluza-Klein monopoles.
These differences might be explained in terms of the Type IIA counterpart
of the configuration; their configuration is
the one in which D6-branes would be dynamical as D4 and NS5-branes rather 
than background.

It is important to comment on the case of $\Nf = 2 \Nc$. In this case,
the D5 charge vanishes, and ${\tilde b^{{\rm NS}}}$ is the constant, which
leads to the gauge coupling constant determined 
by the constant ${\tilde b^{{\rm NS}}}/\gs$. The solution reduces to
\eqn
&& \hspace*{0.5cm} d s^2_{12} = ds^2_{10} 
 +  \left(2\pi \Ap\right)^2\left|d V_2\right|^2
\nn              
&& \hspace*{1cm}\left({2\pi \Ap}\right)^{-1} d s^2_{10}  
  = \fr{\rho^2}{(4\gs\Nc {\tilde b^{{\rm NS}}})^{1/2}} 
      \left\{\eta_{\mu \nu} dx^{\mu} dx^{\nu}\right\}
          + \fr{(4\gs\Nc {\tilde b^{{\rm NS}}})^{1/2}}{\rho^2} 
   \left\{|dV_1|^2 + 
             d {\cal M}_{{\rm R^4/Z_2}}^2 \right\} , \nn
&& \hspace*{1cm} C_{(4)} = \fr{\rho^4}{4\gs\Nc {\tilde b^{{\rm NS}}}}  
    dx^{0} \wedge dx^{1}
          \wedge   dx^{2} \wedge dx^{3} . \hsp{1cm} 
\no
\enn
We have to be careful of the region where
the description of the supergravity will be correct.
We keep the ratio $1/g_{YM}^2 = {\tilde b^{{\rm NS}}}/\gs$ fixed 
with $\gs \ra 0$ and
${\tilde b^{{\rm NS}}} \ra 0$. In addition to that, we also have to
consider the region,  
\eq
\gs {\rm Q_{D3}} = 
  \fr{\gs \Nc}{2\pi} {\tilde b^{{\rm NS}}} \gg 1. 
\en
This means that we need the 
large $\Nc$ limit in the same way as 
the other known SUGRA solutions.\footnote{ 
We also have to keep the ratio ${\Nf}/{\Nc}$ fixed.} 

\begin{figure}
\epsfysize=4.0cm 
\centerline{\epsfbox{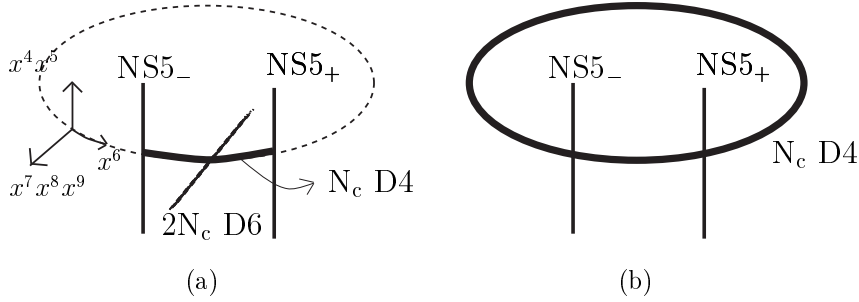}}
\caption{{\small Two conformal configurations: 
(a) partially wrapping $\Nc$ D4-branes  (b) completely wrapping 
$\Nc$ D4-branes}}
\label{wind}
\end{figure}

Note that except for $\rho \sim |V_1|$ in our case, the above solution 
is similar to that of \cite{9803015}.
But, our expression of the D3-brane charge
has ${\tilde b^{{\rm NS}}}$ dependence, but theirs does not. 
The physical meaning of this difference can be explained as the following.
Their configuration consists 
of $\Nc$ D5-branes with 
$({\Nc}b^{{\rm NS}})/{2\pi}$ D3-brane charge
and $\Nc$ anti-D5-branes with 
$\left(2\pi-b^{{\rm NS}}\right){\Nc}/{2\pi}$ D3-brane charge.
As a result of that, the total D3-brane charge is $\Nc$ with 
vanishing D5 charge.
The dependence on $b^{{\rm NS}}$
is gone. 
We can understand this difference more clearly in the 
MQCD configuration as depicted in Fig.\ref{wind} \footnote{
For the comparison of the two cases, the Figure \ref{wind} is depicted
by the same coordinates 
in the region $|u| \gg \Nf\La=2\Nc \La$, although AdS/CFT correspondence is
not applicable in this region, but the perturbative analysis of the field theory
is the good description.}
In their case, the $\Nc$ D4-branes
are wrapping on the circle completely, but in our case they
are wrapping on only a part of the circle.
 
\section{Speculations on the Nonperturbative Effect}

In the previous sections, we have studied in the region where
we can ignore the strong coupling effect $-$ nonperturbative effect.
Let us consider what will happen beyond this perturbative region.  Of
course, we can not extend our analysis to this region, so we have to
limit our discussion within speculation, but this kind of
speculation will be useful.

For example, let us remember MQCD
suggested by \cite{9703166}. This is well known to be the most successful
example in taking in the nonperturbative effect.  The success of MQCD is
based on the fact that in this model the D0-brane is
responsible for the nonperturbative effect (instanton effect) of the 4D
${\cal N}$=2 SQCD. So lifting the whole 
system to the eleven dimensional supergravity
gives the way to take in this effect. Note that in the system in which
D0-brane does not play this key role, lifting to the 11D SUGRA does
not solve the problem automatically.\footnote{For example, let us consider the
NS5(12345)-D2(16)-NS5(12345) system which is T-dual (${\rm T}_{23}$) of
the MQCD configuration $\{$NS5(12345)-D4(1236)-NS5(12345)$\}$. The
nonperturbative effect of this system is due to the 
D2-brane, not D0-brane. This makes
it impossible to take in all the nonperturbative effect only by lifting
to 11D SUGRA. That is, when we lift this system to 11D SUGRA, we can
distinguish the M2-branes from the two M5-brane on which the M2-branes
are ended. This means that we can see the magnitude for the gauge coupling
constant of 2D SYM on the D2-branes and that the dimensional transmutation
does not happen yet.}

Imagine that we did not know the fact that 11D SUGRA includes all the
effects of the D0-brane in Type IIA theory. 
As long as we know that the D0-brane is responsible for
the 4D instanton effect, we could say at least
the followings; if all the effects of the
D0-brane are included, D4-brane and NS5-brane would become the same
thing. This expectation comes from the property of the 4D field theory
that the nonperturbative effect makes
the gauge coupling constant invisible after 
the dimensional transmutation. Moreover, from the knowledge of the
purely field theoretical analysis, we can tell what this
configuration would be like $-$ the
configuration described by the Seiberg-Witten curve.

As seen in this case,
the knowledge about well-known results of the field theory
may enable us to give some clues about the unknown aspects of
the string theory.

So let us speculate what will happen in the model that we have discussed.
What is responsible for the nonperturbative effect in Type IIB theory
?  By the T-duality of the D0-branes in the MQCD configuration, 
we can easily find out that it is the
D1-branes that play that role. These D1-branes are
wrapping on the vanishing 2-cycle on the orbifold.  This is
also confirmed by the analysis of the action \cite{0206063} as done in
\cite{9708041} \cite{9803140} for MQCD configuration.
So it is plausible that the nonperturbative effect would be included
if we could add all the D1-brane effects to the previous result.
But it is technically very difficult to carry out such a task directly. So we have
to limit our discussion within qualitative speculations about the
configuration which would be described by the
Seiberg-Witten curve.\footnote{Here we limit
our discussion within the study of  
the configuration and flux, not gravity solution.
In \cite{0110109}, they have discussed the gravity dual 
for 3D SYM in which the singularity is removed
by adding the 'non-perturbative gauge fields'.}  
But this can be done without using the explicit (direct) calculations 
of the D1-brane effects.  

First, let us consider simple pure SYM theory. In the weak coupling
region, this is the Type IIB configuration 
discussed in \cite{9911096}\cite{0011077}
that is also the case of $\Nf =0$ in our model. In this region, the flow
of the gauge coupling constant is described as the complex field $\gamma \equ
\tau b^{{\rm NS}} - b^{{\rm R}}$.\footnote{Here, we consider the general
cases with $C_0 \neq 0$.}

From the success of MQCD, we know how this complex field behaves. Because
this field corresponds to the distance between the two NS5-branes on the
two-dimensional space $(x^6,x^{10})$ in Type IIA theory, we can get
the exact behavior of this complex field from the Seiberg-Witten curve as
\eq
\gamma\left(u\right) = i (\ln y_+ - \ln y_-) , \hsp{1cm} 
y_{\pm} \equ \sum_{n=0}^{\Nc} s_n u^n \pm \sqrt{\left(\sum s_n u^n\right)^2 
     - \Laq^{2\Nc}}  \ \ . 
\label{gamma}
\en
Here $\{s_n\}$ are the moduli parameters which satisfies the
conditions,  $s_{\Nc} =1$ and $ s_{\Nc-1}=0$. 
We also denote the dynamical scale for this gauge theory
as $\Laq$.
The above $y_{\pm}$ are the
solution of the quadratic equation (Seiberg-Witten curve),
$y + \Laq^{2\Nc} /y = 2\sum_{n=0}^{\Nc} s_n u^n$. \footnote{
In MQCD, the complex coordinate $y$ corresponds
to the real coordinate $(x^6,x^{10})$ by the relation
$\ln y ={(x^6+i x^{10})/R_{10}} = (\ph + i\chi)/2$ in our notation. }

Note that the real
and imaginary parts of $\gamma\left(u\right)$ have the origin of RR and
NSNS 2-form gauge field respectively, as seen in the previous sections. 
But we have only D5-branes
and no NS5-branes in our model. So it is plausible that even in the strong
coupling region, we will obtain a real integer corresponding to the quantized D5
charge and no NS5 charge.

To find out what happens in the strong coupling
region, let study the complex field strength 
\eq
\pa_u \gamma\left(u\right) = 2i
\left(\sum n s_n u^{n-1}\right) \left\{\left(\sum s_n u^n\right)^2 -
\Laq^{2\Nc}\right\}^{-1/2} .
\label{strength}
\en 
As seen in the form of this field strength, there are branch cuts
between the two points, say, $u_{\pm}^{(i)}$
 $(i=1,2, \ldots \Nc)$  
which satisfy $\sum s_n (u_{\pm}^{(i)})^n =\pm \Laq^{\Nc}$. 
They reduce to $u_{+}^{(i)}
=u_{-}^{(i)}$ under the
condition of $\Laq =0$. 
When we integrate the field strength (\ref{strength}) 
around the pairs of these points, 
we obtain the expected result $-$ no NS5 charge and 
the quantized D5 charge
$\fr{1}{2\pi} \oint \pa_u \gamma =2m$.
Here $m$ is the number of the pairs of the branch points 
$(u_{-}^{(i)}, u_{+}^{(i)})$ surrounded 
by the pass of the integral. This reproduces 
the classical picture that
the D5-branes are located at the points which satisfy 
$\sum s_n u^n =0$.
So we can conclude that the 
nonperturbative effect in Type IIB theory causes the splits 
of these classical $\Nc$ positions of the D5-branes.
This is the well-known phenomena in the 4D N=2 gauge theories.
This has a lot of implication $-$ This shows that we can not
exactly tell where the D5-branes are located. They seems to spread on the
$u$-plane and make the different type of singularity from the 
point-like source $-$ branch cut. This corresponds to the situation in MQCD
that the D4-branes become indistinctive of NS5-branes after the
strong coupling effect. On the field theory side, this is
the manifestation of the dimensional transformation.
\begin{figure}
\epsfysize=5.0cm 
\centerline{\epsfbox{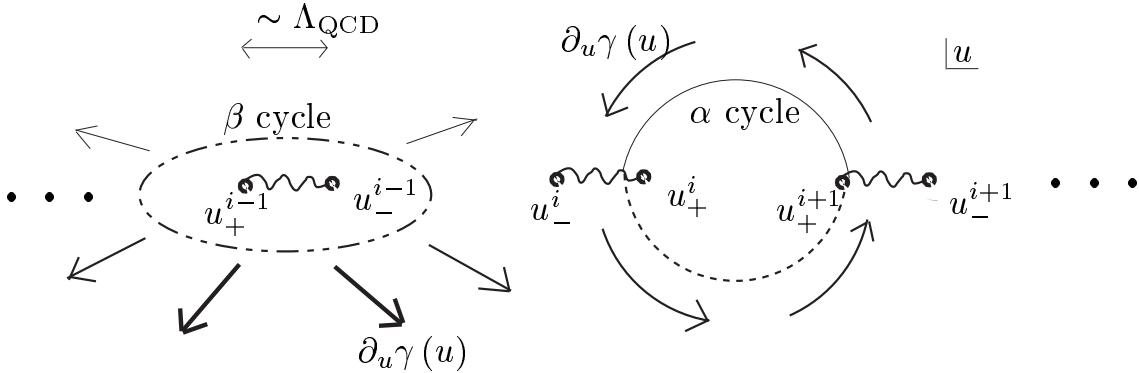}}
\caption{{\small The two kinds of cycles and the 
behavior of the field strength 
$\pa_u \gamma\left(u\right)$.}}
\label{cycle}
\end{figure}

In addition to the above type of the integral pass ($\B$-cycle),
there is also another type of the pass of 
the integral, called $\A$-cycle. This is the pass
which runs around the points, say $(u_{+}^{(i)}, u_{+}^{(i+1)})$ 
crossing the $i$th and $(i+1)$th branch cuts. (See Fig.\ref{cycle})
By this $\A$-cycle integral for the field strength (\ref{strength}),
we can easily see 
$\fr{1}{2\pi} \oint_{\A} \pa_u \gamma =0$.
This means that there is no source from which the flux
goes out. It leads to that the fluxes go around the $\A$-cycles 
from one branch cut to another branch cut. 
  
Can we put the geometrical meaning on this $\gamma\left(u\right)$ ?  We
can interpret that this complex function expresses the point
$(b^{{\rm NS}}, -b^{{\rm R}})$ on the torus with constant complex
structure $\tau =i/\gs + C_0$. In our analysis, we have to limit within
the region, $\gs \ll 1$ and $b^{{\rm NS}} \ll 1$ with the arbitrary
magnitude of the ratio, $b^{{\rm NS}} / \gs$. As a result of that, one
of the two periods of this torus is finite and the other is
infinite. This leads to the conclusion that this complex function
$\gamma\left(u\right)$ expresses the arbitrary point in the
belt-like two-dimensional plane with the topology \mbox{\boldmath ${\rm R} 
\times {\rm S^1}$}.
This also means that we do not have the complete invariance under the
SL(2 Z) transformation $-$ there are the invariance under
T-transformation which is originated from the periodicity of $x_{10}
$-direction in M theory, but no invariance under S-transformation.
\footnote{This fact is also easily confirmed by the observation
as the following;
in corresponding MQCD configuration, 
we have to set the radius of $x^6$-direction infinite in order to
avoid the NS5-branes crossing each other.
An exception 
with S-invariance is the case
with conformal invariance known as the elliptic model in which
NS5-branes are straight without crossing each other.}
Of course, we can directly see this fact from the form of 
this complex function. 

We comment here on the Seiberg-Witten 1-form. This is written as
$\la_{{\rm SW}} \equiv ud\gamma $ and gives us the 
exact expression for the effective gauge coupling constants of 
the low energy U(1)$^{\Nc -1}$ gauge
theory of the 4D
${\cal N}=2$ SU($\Nc$) gauge theory.
The U(1) effective gauge coupling constant 
(perturbatively) corresponds to the 
value of the field $\gs^{-1}b^{{\rm NS}}$ at the point where each D5-brane
is located. As seen in our discussion, this also gives
the expression for the D3-charge induced in each D5-brane. 
So we can expect the exact result for the effective coupling constant
will also give us the exact expression for the D3-brane charge. 
This is also the same in the case with fundamental matters if we
replace $b^{{\rm NS}}$ with ${\tilde b^{{\rm NS}}}$.  
The calculation of these effective gauge coupling constants has 
been done a lot, so
we do not repeat this analysis here. We limit our discussion 
within the comment on this. 

In summary, the non-perturbative (D1-brane) effect will 
be speculated as below:
\ite
\item The classical $\de$ function-like 
singularities as the source of the D5 charge
change into those of the branch cuts.
\item There is the new type of 'flux' \footnote{
The quotation marks are added to mean that this is the flux
after taking in the nonperturbative D-string effect.}
which goes 
round between one branch cut 
and another branch cut. 
\itn
\vsp{0.5cm}

Next, let us consider the case including the (massless) 
$\Nf$ fundamental matters. \footnote{We limit our analysis
in the region $\Nf < 2 \Nc $ in which the gauge
theory is asymptotically free.}
This is almost the same as pure Yang-Mills case except that
the Seiberg-Witten curve is different. This difference leads to the 
modification of $\gamma$ as
\eq
\gamma\left(u\right) = i (\ln y_+ - \ln y_-) , \hsp{1cm} 
y_{\pm} \equ \sum_{n=0}^{\Nc} s_n u^n \pm \sqrt{\left(\sum s_n u^n\right)^2 
     - \Laq^{2\Nc -\Nf} u^{\Nf}}  \ \ .
\label{gammamat}
\en
In the above, $y_{\pm}$ are the
solution of the quadratic equation (Seiberg-Witten curve),
$y + \Laq^{2\Nc -\Nf} u^{\Nf} /y = 2\sum_{n=0}^{\Nc} s_n u^n$. 
Note that $y$ is not the same coordinate as that of pure SYM theory,
but the same as that appeared in our analysis of the previous
sections. \footnote{
The definition of $y$ is given in Eq.(\ref{a_6chi_sol})
and the relation with $a$ and the NSNS 2-form field
is given in Eq.(\ref{adef}).}

Let us rewrite the expression for $\gamma$ as
\eq
\gamma\left(u\right)= i \left\{2\ln \left(\fr{
\sum_{n=0}^{\Nc} s_n u^n + \sqrt{\left(\sum s_n u^n\right)^2 
     - \Laq^{2\Nc -\Nf} u^{\Nf}}}{\Laq^{\Nc}}\right)
-  \Nf \ln \left(\fr{u}{\Laq}\right)\right\}.
\label{matg}
\en
It is easy to see from the first term 
that there are $\Nc$ singularities of the branch cuts.\footnote{
There are the multiple $\Nf$ branch points, but we can resolve this
singularity by giving the mass term.}
Roughly speaking, this shows that classical $\de$ function-like 
singularities of the external $\Nc$ D5-brane source 
changes into those of the branch cuts. 
The second statement in pure SYM theory 
about the two kinds of flux is also applicable to this case with matters. 
But we have to be careful of the second term in the above expression
for $\gamma\left(u\right)$. This gives additional contribution of $-\Nf$ 
D5 charge to the contour integral around the origin. 
So we can roughly say that this term is the 
contribution of the matters or the background, as compared with the first term.
In fact, in the region $|u| \gg \Laq$, we can see the behavior of 
$\gamma\left(u\right)$ with the vanishing moduli $\{s_n\}=0$, as
\eq
\gamma\left(u\right) \sim  i \left(2 \Nc - \Nf \right) 
\ln \left(\fr{u}{\Laq}\right),   
\label{pert}
\en
where the first term in the above comes from the first 
term of Eq.(\ref{matg}). 
This is the perturbative RG-flow in the ultraviolet region 
in the 4D field theory.

Note that we can also obtain this result in the gentler 
region $|u| / \Laq  > 1$ by the large $\Nc$ and $\Nf$
limit. This is the RG-flow in the region where AdS/CFT correspondence
is effective as discussed in section 6.

Therefore as long as
one of the above conditions is satisfied, 
our result for the complex field in the previous sections
is trustworthy.\footnote{Our 
approximation about the source as the heavy
bifundamental quark is justified in this region.} 
\section*{Acknowledgments}
This work is supported  in part by the Japan Society for the
Promotion of Science under the Postdoctoral Research Program (No. 12-08617)

\section*{Appendix}
In this appendix, we will show our analysis in the section 3.1 and 3.2
is the same when we start from the Born-Infeld action.

After taking the static gauge $\sig_0 =t$ and
$\{ \sig_4,\sig_5 \} = \{x_4, x_5\}$, and assuming 
that only $X^{6}=X^{6}(x_4, x_5)$ and $A_0 =A_0(x_4, x_5)$ are
the nontrivial fields,
let us take
the limit (\ref{limit}). Then the action (\ref{BI}) reduces to 
\eqn
S_{{\rm D2}}^{{\rm BI}} &=& - \fr{1}{4\pi \La} \int dt du^4 du^5 
   {\rm H} \left\{{\rm G}\left(\vn_u \ph,\vn_u a\right)\right\}^{1/2}
  + \fr{\La^2}{4\pi}\int dt du^4 du^5 
          \fr{\Nf}{R^2} \left(\fr{\ph}{R} - 
               \vn_u \ph \cdot \fr{\vec{u}}{R}\right)a ,\nn
&&{\rm G}\equiv {\rm G}\left(\vn_u \ph, \vn_u a\right)
        \equ 1 + \La^2\left\{\left|\vn_u \ph \right|^2 
            - \left|\vn_u a \right|^2 \right\}
              - \La^4 \left(\vn_u \ph \times \vn_u a\right)^2 
\label{D2actionBI}
\enn
where we use the same convention in the section 3.1.
Let us add the source with $\pm \Nc$ electric charges
to the above action. From the action 
$S_{{\rm D2}}^{{\rm BI}}  + \De S $, we can see
the constraint for $a$ (Gauss-Low), 
\eqn 
&&\hspace*{-1cm}\vn_u \left({\rm H G^{\fr{-1}{2}}}\vn_u a\right) 
  + \La^2 \vn_u \times \left\{{\rm H G^{\fr{-1}{2}}}
     \vn_u \ph \left(\vn_u a \times \vn_u \ph\right)\right\}
= \fr{\Nf \La}{R^3}
\left(\ph - \vn_u \ph \cdot \vec{u} \right) 
\pm 4\pi \Nc \de (u^4)\de (u^5)  \nn
&&\hspace*{6cm}= \vn_u \left(-\fr{\Nf \La
\ph}{R} \fr{\vec{u}}{|u|^2} + \left\{{\rm sign} (\ph_0) \Nf \pm 
2\Nc  
\right\}\fr{\vec{u}}{|u|^2}
\right), 
\label{GaussBI}
\enn 
where the right hand is the same as that of the section 3.1.
Next, let us consider the equation of motion. We can easily obtain
\eqn 
\hspace*{-1cm} &&\vn_u \left({\rm H G^{\fr{-1}{2}}}\vn_u \ph\right) 
  - \La^2 \vn_u \times \left\{{\rm H G^{\fr{-1}{2}}}
     \vn_u a \left(\vn_u \ph \times \vn_u a\right)\right\}
=  - \fr{\Nf \La }{R^3} \left({\rm G^{\fr{-1}{2}}}\ph 
             + \vn_u a \cdot \vec{u}\right)
               \mp 4\pi \Nc  \de^2({\vec u}).
\no
\enn 
From this form, we can see that the same additional
relation $\vn a = - \vn \ph$ which appears in the section 3.1.
make the above equation equivalent to the Gauss-Low (\ref{GaussBI}). 
As a result of that, we obtain the same equation as Eq.(\ref{ph_eq})
to determine the behavior of $\ph$ as that of the section 3.1.

Then let us discuss the field $\chi$.
The electric charge 
${\rm Q_E}$ in this case is given by the integral of the left hand as     
\eqn
&&\hspace*{-0.5cm}{\rm Q_E} = \fr{1}{4\pi} \oint *\left\{
{\rm H G^{\fr{-1}{2}}}\vn_u a
   + \La^2  {\rm H G^{\fr{-1}{2}}} \tvn_u \ph  
         \left(\vn \ph \times \vn a\right)
   + \Nf\left(\fr{\La \ph}{R}  
   - {\rm sign} (\ph_0)\right)\fr{{\vec u}}{|u|^2}
\right\}
 =  \pm \Nc, \no
\enn
By taking the current of $\ph_0 > 0$ 
as the standard. we define the 'dual' field $\chi$ as 
\eq 
\tvn_u \chi \equ 
{\rm H G^{\fr{-1}{2}}}\vn_u a
   + \La^2  {\rm H G^{\fr{-1}{2}}} \tvn_u \ph  
         \left(\vn \ph \times \vn a\right)
+ \Nf\left(\fr{\La \ph}{R}  
   - 1\right)\fr{{\vec u}}{|u|^2}
\label{chiBI}
\en
By this definition, we can see the solution for $\chi$ is the same as
(\ref{chi}).

In summary, the final result for $\phi$ and $\chi$ are the same as those
of the section 3.1 
even when we start with the Born-Infeld action (\ref{D2actionBI}).  
\newcommand{\NP}[1]{Nucl.\ Phys.\ {\bf #1}}
\newcommand{\AP}[1]{Ann.\ Phys.\ {\bf #1}}
\newcommand{\PL}[1]{Phys.\ Lett.\ {\bf #1}}
\newcommand{\NC}[1]{Nuovo Cimento {\bf #1}}
\newcommand{\CMP}[1]{Comm.\ Math.\ Phys.\ {\bf #1}}
\newcommand{\PR}[1]{Phys.\ Rev.\ {\bf #1}}
\newcommand{\PRL}[1]{Phys.\ Rev.\ Lett.\ {\bf #1}}
\newcommand{\PRE}[1]{Phys.\ Rep.\ {\bf #1}}
\newcommand{\PTP}[1]{Prog.\ Theor.\ Phys.\ {\bf #1}}
\newcommand{\PTPS}[1]{Prog.\ Theor.\ Phys.\ Suppl.\ {\bf #1}}
\newcommand{\MPL}[1]{Mod.\ Phys.\ Lett.\ {\bf #1}}
\newcommand{\IJMP}[1]{Int.\ Jour.\ Mod.\ Phys.\ {\bf #1}}
\newcommand{\JP}[1]{Jour.\ Phys.\ {\bf #1}}
\newcommand{\JAG}[1]{Jour.\ Alg.\ Geom.\ {\bf #1}}

\end{document}